\documentclass[aps,superscriptaddress,twocolumn,]{revtex4}
\usepackage{bm}
\usepackage{epsfig}
\usepackage{times}
\usepackage{bbm}
\usepackage{amssymb}
\usepackage{amsmath}
\usepackage{comment}
\usepackage{natbib}
\usepackage{color}
\usepackage{graphicx}
\usepackage[normalem]{ulem} 
\usepackage[latin1]{inputenc}
\usepackage[loose,nice]{units}       

\begin{document}
\title{Absorption and analysis of unbound quantum particles -- one by one}

\author{S{\o}lve Selst{\o}}
\affiliation{Faculty of Technology, Art and Design, Oslo Metropolitan University, NO-0130 Oslo, Norway}

\begin{abstract}
In quantum physics, the theoretical study of unbound many-body systems is typically quite demanding -- owing to the combination of their
large spatial extension and the so-called {\it curse of dimensionality}.
Often, such systems are studied on truncated numerical domains -- at the cost of information.
Here we present methods for calculating differential probabilities for unbound particles
which are subject to a {\it complex absorbing potential}.
In addition to attenuating outgoing waves, the absorber is also used to {\it probe} them by
projection onto single-particle scattering states,
thus rendering the calculation of multi-particle scattering states superfluous.
Within formalism based on the Lindblad equation, singly differential spectra from subsequent absorptions are obtained by resolving the dynamics of the remaining particles after the first absorption. While the framework generalizes naturally to any number of particles, explicit, compact and intuitive expressions for the differential probability distributions are derived for the two-particle case. The applicability of the method is illustrated by numerical examples involving two-particle model-systems.
These examples, which address scattering and photo ionization, demonstrate how energy distributions of unbound particles may be determined on numerical domains considerably smaller than the actual extension of the system.
\end{abstract}

\maketitle

\section{Introduction}
\label{Introduction}

Simulating quantum many-body dynamics is often a challenging endeavour.
One reason for this is the fact that the complexity of such studies grows exponentially with the number of particles; this is the infamous {\it curse of dimensionality}. For unbound many-body quantum systems it becomes even worse as the extension of the systems under study is not limited. The numerical study of such systems requires a very high number of degrees of freedom -- for each particle. Combined with the curse of dimensionality, this renders several interesting simulations unfeasible.

A much applied way of dealing with the unbounded nature of the wave function is to impose absorbing boundary conditions. Such boundary conditions allow us to remove outgoing waves, corresponding to unbound particles, and truncate the numerical domain without introducing artifacts such as reflections at the boundary or wave packets reappearing at the opposite edge of the grid, as would be the case with periodic boundary conditions. Absorbing boundaries may be introduced in several ways \cite{DeGiovannini2015}; {\it exterior complex scaling} \cite{McCurdy2004} or the closely related notion of {\it perfectly matched layers} \cite{Scrinzi2014} are frequently used techniques. Another one is to introduce a masking function in the propagation scheme \cite{Krause1992,Chelkowski1998,Grobe1999,Lein2002}, or, equivalently, to augment the Hamiltonian with a complex absorbing potential,
a {\it CAP},
which vanishes in some interior region \cite{Weisskopf1930,Kosloff1986,Scrinzi1981,Riss1993,Rescigno1997,Moiseyev1998,Manolopoulos2002,Santra2002,Muga2004,Sajeev2006}. Such a potential could, like exterior complex scaling, depend on both position and momentum, or it could be a purely position-dependent potential. In this work, we will exclusively deal with the latter.

When a simulation of a dynamical quantum systems is subjected to absorption, information is lost during the course. In a many-body setting, this loss is devastating for simulations based on the Schr{\"o}dinger equation alone; if one particle is absorbed, the entire wave function is lost. No information remains about the remaining sub-system. While this issue is resolved by the Lindblad equation \cite{Lindblad1976,Gorini1976}, the resulting equation of motion is still such that the information about the absorbed particles is discarded. However, since we know precisely what is removed from one instant to the next, it should be possible to analyse this removed part as it is absorbed. By aggregating such contributions from each time step, information such as, e.g., the energy spectrum of the unbound particles should be obtainable even on a truncated numerical grid. This is what we aim to do in this paper.

This is, of course, by no means any new endeavor. Several interesting methods for obtaining energy or angle resolved information about unbound particles using truncated numerical grids
are put forward, see, e.g., \cite{Chelkowski1998,Ermolaev1999,Ermolaev2000,Beck2000,Feuerstein2003, Palacios2007,Greenman2010,Tao2012,Scrinzi2012,Yip2013,Serov2013,
Karamatskou2014,Majety2015_1,Morales2016,Wang2018}.
These methods have enabled several interesting physical studies within atomic, molecular and optical physics, see, e.g., \cite{Serov2001,Tong2006,Palacios2008,Horner2008,Rohringer2009,Palacios2009,Palacios2010,Liertzer2012,Argenti2013,Yue2013,Granados2013,
Yue2014,Yip2015,Majety2015_2,Zielinski2016,Pont2016,Majety2017} -- only to name a few.

Many of these schemes are formulated in a one-particle context -- with no obvious generalization to a many-body context. However, some of them have 
been adapted to formulations involving Hartree-Fock orbitals with a limited number of excitations~\cite{Beck2000,Greenman2010,Karamatskou2014,Majety2015_1}. The methods of \cite{Palacios2007,Scrinzi2012} are noteworthy exceptions, however, as they deal explicitly with two-particle systems in a general framework. In the former, exterior complex scaling is imposed in order to analyze the wave function immediately after interaction. The need for calculating correlated scattering states explicitly is circumvented by solving a set of time-independent inhomogeneous Schr{\"o}dinger equations. In this context, exterior complex scaling is used to impose outgoing boundary conditions; it is not directly involved in the dynamical calculations. In Ref.~\cite{Scrinzi2012}, on the other hand, absorbers are used to attenuate outgoing waves for a two-electron atom while the interaction with a laser pulse is still ongoing. It demonstrates how differential information may be obtained by monitoring the flux through a surface in the asymptotic region and reconstructing an approximate wave function in terms of Volkov states. This approximation is valid when the surface and absorber is placed in the asymptotic region in which the Coulomb interaction with nucleus the electron-electron repulsion may be neglected.

The method presented here follows a rather different path. It takes the proper formulation of particle loss due to an absorber as the starting point. This implies a departure from a pure wave-function description. Beyond the introduction of a CAP, it does not resort to any approximation, {\it ansatz} or model, and probability is conserved manifestly without the introduction of any heuristic arguments. 
It involves projections onto single-particle scattering states only -- irrespective of the initial number of particles involved. This introduces a significant reduction in complexity since it evades the need for calculating many-particle scattering states, which in general involve several, possibly multiple,
continua.

The ability to analyse the unbound particles using single-particle scattering states relies on the fact that the absorber is a one-particle operator. It only removes one particle at a time, and no two-particle interaction is involved in the absorption process. 
The CAP is used actively to {\it probe} the outgoing waves.
We will demonstrate how differential information about unbound particles can be obtained by accumulating information about the particles which are absorbed -- one by one.
Our approach provides a generic and intuitive scheme which generalizes naturally to any number of particles.
Moreover, analyzing the outgoing waves introduces little extra effort in terms of implementation, computation and memory requirements.
A drawback may be that it produces a set of singly differential probability distributions only, no fully differential distributions for several particles simultaneously.

In the next section, the method is explained in detail.
Explicit formulas are derived for the two-particle case. In Sec.~\ref{Sec_NumExamples}, we illustrate the scheme by applying it to two specific model systems, one involving scattering and another involving photo ionization. Conclusions are drawn in Sec.~\ref{Sec_Conclusions}.

\section{Theory}
\label{Sec_Theory}

Our starting point is a quantum system consisting of $N$ identical particles. We impose absorbing boundaries by adding a CAP to the Hamiltonian.
The resulting effective Hamiltonian is then
\begin{equation}
\label{EffectiveHam}
H_\mathrm{eff} = H - i \Gamma ,
\end{equation}
where the actual Hamiltonian, $H$, and the CAP, $\Gamma$, are both Hermitian. Moreover, $\Gamma$ is positive semi-definite.
In a many-body context, it is convenient to express the interactions in terms of second quantization; this allows us to write up $H$ and $\Gamma$ in a manner which does not explicitly depend on the number of particles. We assume that $H$ contains interactions between at most two-particles, while the CAP, $\Gamma$, is a one-particle operator. With a local potential the CAP is diagonal in position representation. Specifically,
\begin{equation}
\label{DefGamma}
\Gamma = \int \gamma(x) \hat{\psi}^\dagger(x) \hat{\psi}(x) \, dx,
\end{equation}
where $\hat{\psi}(x)$ annihilates a particle with the position coordinates of $x$ while $\hat{\psi}^\dagger(x)$ creates a particle with coordinates of $x$. For identical fermions these field operators obey the usual anti-commutation rule
\begin{equation}
\label{AntiCommutator}
\left\{ \hat{\psi}(x), \hat{\psi}^\dagger(x') \right\} = \delta(x-x') ,
\end{equation}
where the anti-commutator is replaced by a commutator in the case of identical bosons. Here ``$x$'' is taken to mean {\it all} degrees of freedom, including position, for each particle, and ``$\int \, \boldsymbol{\cdot} \, dx$'' refers to the definite integral or sum over the entire domain.

The CAP function $\gamma(x)$ is assumed to be zero within some finite interaction region and positive beyond. We will, as mentioned, take it to depend exclusively on the spatial coordinates.
It is usually imposed
in order to avoid artifacts such as reflections at the boundary. In case of too high $\gamma(x)$-values, hard absorption may still induce reflections, however.
As the CAP is not introduced on physical grounds, any dependence on $\Gamma$ in the results of numerical simulations is unphysical. Results of simulations which prevail in the limit that $\Gamma$ vanishes, may, however, be considered physical and correct.

\subsection{Absorption from an $N$-particle system}
\label{Sec_AbsFromN}

The evolution of an $N$-particle system subject to a CAP is governed by a non-Hermitian Schr{\"o}dinger equation,
\begin{equation}
\label{TDSE}
i \hbar \frac{d}{dt} | \Psi_N (t) \rangle  = H_\mathrm{eff} | \Psi_N (t) \rangle .
\end{equation}
In going from time $t$ to $t+\tau$, the state evolves into
\begin{equation}
\label{EvolutionPureState}
| \Psi_N(t+\tau) \rangle = | \Psi_N (t) \rangle - i \frac{\tau}{\hbar} H | \Psi_N (t) \rangle - \frac{\tau}{\hbar} \Gamma | \Psi_N(t) \rangle 
\end{equation}
to leading order in $\tau$.
The last term above leads to depletion of the original $N$-particle system; the part $-\tau/\hbar \, \Gamma | \Psi_N(t) \rangle$ has been removed in this time step. Since we know what we remove, we can also {\it analyse} this removed part. Specifically, one could project this part onto the appropriate scattering states and accumulate their contributions over time. Naively, this would correspond to accumulating contributions of the form
\begin{equation}
\label{NaiveProjection}
\left| \left \langle \varphi^N_\varepsilon \right| \frac{\tau}{\hbar} \Gamma \left | \Psi_N \right \rangle \right|^2 ,
\end{equation}
where ``$\varepsilon$'' specifies the set of physical quantities of interest for the $N$-particle scattering state $|\varphi^N_\varepsilon \rangle$.
This is indeed naive since such contributions, being of order $\tau^2$, would not integrate to the total probability of absorption.
If one, alternatively,
tries to integrate such contributions in time in a coherent manner, it is still not obvious how such a formulation would be consistent with norm conservation. 

Instead, it is more instructive to consider the density matrix $\rho_N = | \Psi_N \rangle \langle \Psi_N |$. The evolution of Eq.~(\ref{TDSE}) may equivalently be described by the non-Hermitian von Neumann equation:
\begin{equation}
\label{TDvNE}
i \hbar \frac{d}{dt} \rho_N = H_\mathrm{eff} \rho_N - \rho_N H_\mathrm{eff}^\dagger =
[ H, \rho_N ] - i\{ \Gamma , \rho_N \} .
\end{equation}
Now, in going from time $t$ to $t+\tau$, the part which has been removed from the density matrix is, again, to leading order
\begin{equation}
\label{RemovedFromRho}
\frac{\tau}{\hbar} \{ \Gamma, \rho_N \} = \frac{\tau}{\hbar} \big( \,
 \Gamma | \Psi_N \rangle \langle \Psi_N | + | \Psi_N \rangle \langle \Psi_N | \Gamma \, \big) ,
\end{equation}
which contributes
\begin{equation}
\label{ExpectationValueRemoved}
\frac{\tau}{\hbar} \langle \varphi_\varepsilon^N |
\{ \Gamma, \rho_N \} | \varphi_\varepsilon^N \rangle =
\frac{2 \tau}{\hbar} \Re \text{e} \, \langle \varphi_\varepsilon^N | \Gamma | \Psi_N \rangle \langle \Psi_N | \varphi_\varepsilon^N \rangle
\end{equation}
to the $\varepsilon$-differential probability distribution of the absorbed particle. In this formulation, the resulting differential probability indeed provides the total absorption probability when integrated over time and $\varepsilon$.

This could be a viable path for processes in which only one particle is liberated. However, if there is a significant probability of several particles becoming unbound in the process, it is likely to require a CAP which is very weak and/or placed quite far away from the interaction region. The reason for this is the fact that, in order for Eq.~(\ref{ExpectationValueRemoved}) to describe the multi-particle process correctly, all particles must have reached their respective continua before absorption sets in. Suppose we have the extreme opposite situation in which one particle is fully absorbed before the other particles have had a chance to reach their respective continua. Then the wave function, which is normalized to the probability of having $N$ particles on the grid, becomes identically zero. Thus, no information at all remains about the other particles -- regardless of wether any of these would have been liberated at a later time or not. Correspondingly, the multi-particle contributions to Eq.~(\ref{ExpectationValueRemoved}) would be suppressed unless the absorption is delayed until all interaction is over.

In this work we will 
We will base our approach on dynamical equations which allow us to maintain the remainder of the system as one particle undergoes absorption. This cannot be done within any approach based on the Schr{\"o}dinger equation alone. 
As will be explained in Sec.~\ref{Sec_LindbladRevisited}, 
this comes about via the Lindblad equation. 
Moreover, 
we will reformulate the projections in terms of 
single-particle scattering 
states. This, in turn, enables us to 
calculate the differential probability distributions arising from subsequent absorption of multiple particles.

Instead of projection onto $N$-particle scattering states, as in Eq.~(\ref{ExpectationValueRemoved}), we integrate out all degrees of freedom for all but one of the identical particles and then analyze the remaining one.
This may be done
by means of field operators. The time derivative of the singly differential probability distribution, $\partial P/ \partial \varepsilon$, can, accordingly, be expressed as
\begin{align}
\nonumber
& \hbar \frac{d}{dt} \frac{\partial P}{\partial \varepsilon} =
\\
&
\label{ExpectationValueRemovedSinglePart}
\int \int \cdots \int_{N-1}
\langle \varphi_\varepsilon |
\hat{\psi}(x) \hat{\psi}(x') \cdots \hat{\psi}(x^{(N-1)})
\{ \Gamma, \rho_N \}
\\ &
\nonumber
\times \hat{\psi}^\dagger(x^{(N-1)}) \cdots \hat{\psi}^\dagger(x') \hat{\psi}^\dagger(x)
| \varphi_\varepsilon \rangle \, dx \, dx' \cdots dx^{(N-1)} ,
\end{align}
where $| \varphi_\varepsilon \rangle$ now is a single-particle scattering state corresponding to the single physical quantity $\varepsilon$. The choice of naming it ``$\varepsilon$'' reflects the fact that {\it energy} is often the quantity in question; it could, however, be any relevant quantity, vectorial or scalar.

We should also address one issue from which both Eq.~(\ref{ExpectationValueRemoved}) and (\ref{ExpectationValueRemovedSinglePart}) suffer. The fact that both $\Gamma$ and $\rho_N$ are Hermitian, ensures that the anti-commutator $\{ \Gamma, \rho_N\}$ is also Hermitian, and, thus, any diagonal element of this 
anti-commutator 
is real. However, although $\Gamma$ and $\rho_N$ are both positive semi-definite, 
$\{ \Gamma, \rho_N\}$ is not necessarily so.
Thus, diagonal elements such as $\langle \varphi_\varepsilon^N| \{ \Gamma, \rho_N\} | \varphi_\varepsilon^N \rangle$ are not manifestly non-negative, which, in turn, makes their interpretation in terms of probabilities dubious.
However, as mentioned, physical results are to be obtained in the limit that the CAP vanishes. As we will see in the numerical examples in Sec.~\ref{Sec_NumExamples}, the problem of ``negative probabilities'' vanishes in this limit.

\subsection{Differential probability distributions from subsequent absorptions}
\label{Sec_LindbladRevisited}

When one out of $N$ particles is absorbed in a simulation as dictated by the Schr{\"o}dinger equation with a non-Hermitian effective Hamiltonian, Eq.~(\ref{TDSE}), the wave function vanishes. It is certainly not converted into any $(N-1)$-particle wave function. All information about the evolution of the remaining particles is lost.
As mentioned, this problem is remedied by the Lindblad equation. Since absorption is in fact a Markovian process and trace and complete positivity should be conserved, the Lindblad equation is the proper starting point. 
By comparing the Lindblad equation in generic form with Eq.~(\ref{TDvNE}), where the CAP expressed in terms of second quantization, Eq.~(\ref{DefGamma}), we may identify a source term which restores the remainder of the system  while one particle undergoes absorption.
The details on how this comes about  
are provided in \cite{Selsto2010}, while an adaption to the multi-configurational time-dependent Hartree-Fock-method is provided in \cite{Kvaal2011}.

The resulting equations of motion constitute a hierarchy of $n$-particle sub-systems, with $n=N,N-1,\cdots, 1, 0$.
The evolution of the $n$-particle sub-system is governed by the master equation
\begin{equation}
\label{TDLE_nPart}
i \hbar \frac{d}{dt} \rho_n = [H, \rho_n] - i \{\Gamma, \rho_n\} + i \mathcal{S}[\rho_{n+1}] ,
\end{equation}
where the source term
\begin{equation}
\label{SourceTerm}
\mathcal{S}[\rho] = 2 \int \gamma(x) \hat{\psi}(x) \rho \hat{\psi}^\dagger(x) \, dx .
\end{equation}
For an initial $N$-particle state, there is no source term and Eq.~(\ref{TDLE_nPart}) reduces to Eq.~(\ref{TDvNE}), which, for a pure state, is equivalent to thenon-Hermitian Schr{\"o}dinger equation, Eq.~(\ref{TDSE}).
As the first particle undergoes absorption at a certain probability, the source term in Eq.~(\ref{SourceTerm}) restores the other particles by populating an $(N-1)$-particle sub-system in a manner which ensures that the total population of the $N$ and the $(N-1)$-particle remains unity in sum. As yet another particle is absorbed, also the $(N-2)$-particle density matrix is populated -- and so on. This way, one may, e.g., distinguish between single and double ionization probabilities of atoms without resorting to many-body scattering states nor numerical domains extending far into the asymptotic region~\cite{Selsto2011}.

As a particle is removed from the $n$-particle sub-system, $\rho_n$, by the absorber, it may be analyzed completely analogously to Eq.~(\ref{ExpectationValueRemovedSinglePart}) by simply replacing $N$ by $n$ in the equation. In doing so, we must, however, take some care in order to avoid double counting.

In the following, we will explain this in detail for the case of $N=2$. We will also develop explicit formulas for the differential probability distributions corresponding to one and two absorptions from the original two-particle system.

\subsection{The two-particle case}
\label{Sec_TwoPartCase}

With an initial two-particle system, Eq.~(\ref{ExpectationValueRemovedSinglePart}) may be written
in
a rather compact form. In App.~\ref{App_dPdE_TwoPart} it is derived how Eq.~(\ref{ExpectationValueRemovedSinglePart}) with $N=2$ for an initial pure state leads to the following expression for the differential probability distribution:
\begin{align}
\nonumber
& \hbar \frac{d}{dt} \frac{dP}{d \varepsilon} =
\int  \langle \varphi_\varepsilon | \hat{\psi}(x) \{\Gamma, \rho_2 \}  \hat{\psi}^\dagger(x) | \varphi_\varepsilon \rangle \, dx =
\\
\nonumber
&
2 \int \int  \, \varphi_\varepsilon(x)^* \varphi_\varepsilon(x')
(\gamma(x) + \gamma(x'))
\\ &
\label{dPdETwoPart}
\times
 \int dy \, \Psi_2(x,y) \Psi_2^*(x',y) dy \, dx dx'
\\ &
\nonumber
+\ 4 \int \, \gamma(x) \left| \int  \, \varphi_\varepsilon^*(y) \Psi_2(x,y) \, dy
\right|^2 \,  dx .
\end{align}
Here $\Psi_2(x_1,x_2)$ is the two-particle wave function in product state representation, cf. Eq.~(\ref{TwoPartWF}).

Eqs.~(\ref{ExpectationValueRemoved}) and (\ref{ExpectationValueRemovedSinglePart}) were proposed from considering the part of the $N$-particle wave function which was removed in a time-step, cf. Eq.~(\ref{RemovedFromRho}). However, these expressions do not take into account that as one particle is removed, the other particle is, via the source term in Eq.~(\ref{SourceTerm}), restored within the one-particle sub-system. From within this sub-system, which is described by the density matrix $\rho_1$, also the second particle may go on to be absorbed and, thus, contribute to a $\varepsilon$-differential distribution obtained from this second absorption.
Let us label the $\varepsilon$-differential distribution obtained from the first absorption, i.e., the absorption from the original two-particle system, by ``$\partial P_2/\partial \varepsilon$'',  while the possible second absorption, from $\rho_1$, will give rise to a distribution labelled ``$\partial P_1/\partial \varepsilon$''. In the first absorption both the particle undergoing absorption and the one that is transferred to $\rho_1$ will contribute in Eq.~(\ref{dPdETwoPart}). However, only the one actually undergoing absorption {\it should} contribute to $\partial P_2/\partial \varepsilon$; the other one will have a chance to contribute to $\partial P_1/\partial \varepsilon$.
Thus, in order to avoid double counting, we must remove the contribution stemming from the particle which is restored within $\rho_1$.

This is not altered by the fact that the particles are indistinguishable nor by the fact that they may undergo absorption at the same time. Actually, as both particles may overlap with the CAP simultaneously, this is not a mere technicality; numerical investigations show that conservation of probability is indeed violated unless 
the term
\begin{equation}
\langle \varphi_\varepsilon | \mathcal{S}[\rho_N] | \varphi_\varepsilon \rangle
\end{equation}
is 
removed from the right hand side of Eq.~(\ref{dPdETwoPart}).
As explained in App.~\ref{App_dPdE_TwoPart}, this contribution
coincides with the last term in Eq.~(\ref{dPdETwoPart}).

After having removed it, we integrate the remaining expression over time and arrive at the following formula for the $\varepsilon$-differential probability distribution of the first particle to undergo absorption from the two-particle system:
\begin{align}
\label{dPdE_General}
\frac{\partial P_2}{ \partial \varepsilon}
& = \frac{2}{\hbar} \int \int \, \varphi_\varepsilon^*(x) \varphi_\varepsilon(x')
\\
\nonumber
& \times \left\{ \gamma(x) + \gamma(x') \right\} \Phi(x,x') \, dx dx' ,
\end{align}
where
\begin{equation}
\label{PhiDef}
\Phi(x,x') = \int_0^\infty \int \Psi_2(x,y; t) \Psi_2^*(x',y; t) \, dy \, dt .
\end{equation}
Here, $\Phi$ amounts to an effective time-integrated one-particle density matrix. This ``density matrix'' is weighted by the CAP function for each of the two variables. These terms are then added and projected onto the scattering states of interest. 
In the special case that the quantity in question, $\varepsilon$, is position -- or some quantity which is purely position-dependent, the scattering states $\varphi_\varepsilon(x)$ and $\varphi_\varepsilon(x')$ become delta-functions, and the contributions to Eq.~(\ref{dPdE_General}) stemming from different times accumulate in an incoherent manner. Otherwise, this aggregation takes place in a manner which maintains coherence; outgoing waves absorbed at different times are allowed to interfere.

While the wave function $| \Psi_2 \rangle$ 
is attenuated 
as the first particle is absorbed, we may
continue to simulate the dynamics of the remaining one within the 
one-particle sub-system. 
The one-particle density matrix $\rho_1$ follows Eq.~(\ref{TDLE_nPart}) with $n=1$.
As we are dealing with a one-particle system in this case, the diagonal matrix element in Eq.~(\ref{ExpectationValueRemovedSinglePart}) coincides with that of Eq.~(\ref{ExpectationValueRemoved}):
\begin{equation}
\label{dPdEOnePart}
\hbar \frac{d}{dt} \frac{\partial P_1}{\partial \varepsilon} =
\langle \varphi_\varepsilon | \{ \Gamma, \rho_1 \} |\varphi_\varepsilon \rangle .
\end{equation}
With only one remaining particle, nothing needs to be removed in order to avoid double counting. The resulting differential probability for the second absorption event reads
\begin{align}
\label{dPdE_GeneralOne}
& \frac{\partial P_1}{\partial \varepsilon} =
\frac{1}{\hbar}  \int \int \varphi^*_\varepsilon(x) \varphi_\varepsilon(x')
\\ &
\nonumber
\times  \{ \gamma(x) + \gamma(x') \} \int_0^\infty  \rho_1(x,x'; t) \, dt  \,  dx dx' ,
\end{align}
see App.~\ref{App_dPdE_TwoPart} for details.
Note that Eq.~(\ref{dPdE_GeneralOne}) coincides with Eq.~(\ref{dPdE_General}) if we substitute the time-integral of $\rho_1$ with $2 \Phi$.

In Eqs.~(\ref{dPdE_General}) and (\ref{dPdE_GeneralOne}) we have analysed the absorbed wave by projection onto time-independent scattering states. When the quantity in question, $\varepsilon$, bears explicit time-dependence, obtaining converged results may require time-dependent scattering states, in which case Eqs.~(\ref{dPdE_General}, \ref{dPdE_GeneralOne}) must be modified to
\begin{align}
\nonumber
\frac{\partial P_2}{ \partial \varepsilon}
& = \frac{2}{\hbar} \int_0^\infty \int \int \, \varphi_\varepsilon^*(x; t) \varphi_\varepsilon(x'; t)
\left\{ \gamma(x) + \gamma(x') \right\}
\\
\label{dPdE_Time}
& \times
\int \Psi_2(x,y; t) \Psi_2^*(x',y; t) \, dy
dx dx' \, dt.
\end{align}
and
\begin{align}
\label{dPdE_GeneralOne_Time}
& \frac{\partial P_1}{\partial \varepsilon} =
\frac{1}{\hbar}  \int_0^\infty \, \int \int  \varphi^*_\varepsilon(x; t) \varphi_\varepsilon(x'; t)
\\ &
\nonumber
\times  \{ \gamma(x) + \gamma(x') \} \rho_1(x,x'; t) \, dx dx' \, dt ,
\end{align}
respectively.
This is somewhat analogous to what is done within the {\it time-dependent surface flux method} and in the {\it mask method}, in which photo electron spectra are calculated during the interaction with an external electric field by projection onto Volkov states, i.e., the eigen-states of the Hamiltonian for a free particle exposed to electromagnetic radiation, see, e.g., \cite{Chelkowski1998,Ermolaev1999,Tao2012,Scrinzi2012,DeGiovannini2012,Serov2013}.

\section{Numerical examples}
\label{Sec_NumExamples}

We will consider two 
examples here, both of which involve two interacting particles in one-dimension. The first one, which is 
without explicit time dependence in the Hamiltonian, addresses a scattering event which could be realized in a quantum dot with narrow confinement in two orthogonal directions. The second addresses a system trapped in the ground state of a confining potential exposed to a pulse of electromagnetic radiation. It may serve as a model for photo ionization of an atom with two active electrons. 
In both cases, the two-particle Hamiltonian may be written
\begin{equation}
\label{TwoPartHamExample}
H_\mathrm{eff} = h_\mathrm{eff}^{(1)} + h_\mathrm{eff}^{(2)} + W
\end{equation}
where the one-particle Hamiltonians $h_\mathrm{eff}^{(i)}$ contain a time-independent, Hermitian part, $h_0$, the CAP and, possibly, a time-dependent perturbation.
The particles interact via a regularized Coulomb interaction:
\begin{equation}
\label{Interaction}
W(x_{12}) =  \frac{W_0}{\sqrt{x_{12}+s^2}} ,
\end{equation}
where $x_{12}=|x_1-x_2|$ and $s$ is a smoothness parameter. We have chosen to use a square CAP function,
\begin{equation}
\label{CAPfunction}
\gamma(x) = \left\{\begin{array}{lc} \gamma_0 (|x|-x_0)^2, & |x|\geq x_0 \\ 0, & |x|<x_0 \end{array} \right. ,
\end{equation}
where $x_0$ defines the onset of the CAP region.

Here and in the reminder of the paper, ``$x$'' refers exclusively to the position variable, and our dynamical variable of interest, $\varepsilon$, will be energy.
In both examples we will deal with states which are symmetric under exchange of the spatial variables, which corresponds to a spin singlet state for fermions. Thus, these particles are, formally, bosons in this context.

Before presenting numerical results, we briefly outline how the calculations are implemented.

\subsection{The implementation}
\label{Sec_Implementation}

The time evolution of the two-particle system, i.e., the solution of Eq.~(\ref{TDSE}), is obtained by a second order split operator technique~\cite{Feit1982}. The implementation is facilitated by expressing the two particle wave function $\Psi_2(x_1, x_2; t)=\langle x_1, x_2 |\Psi_2 (t) \rangle$, where
$|x_1, x_2 \rangle$ is a product state, in terms of a matrix. This way, the action of the one-particle parts of the Hamiltonian, $h_\mathrm{eff}^{(1)}$ and $h_\mathrm{eff}^{(2)}$, corresponds to left and right multiplication, respectively, while the action of the particle-interaction, $W \Psi_2(t)$, corresponds to elementwise multiplication (Hadamard product).
The momentum operator is represented by means of the fast Fourier transform (FFT).

The evolution of the one-particle sub-system follows Eq.~(\ref{TDLE_nPart}) with $n=1$:
\begin{equation}
\label{TDLE_1part}
i \hbar \frac{d}{dt} \rho_1  =  h_\mathrm{eff} \rho_1 -
\rho_1 h_\mathrm{eff}^\dagger
+ 2i
\mathcal{S}[| \Psi_2 \rangle \langle \Psi_2 |] .
\end{equation}
In our matrix formulation, the last term on the right hand side, the source term, Eq.~(\ref{SourceTerm}), may be calculated as
\begin{equation}
\label{SurceTermMatrix}
S = 4 h \Psi_2(t) D \Psi_2^\dagger(t) ,
\end{equation}
where the diagonal matrix $D=\mathrm{diag}(\gamma(\bf{x}))$ is the CAP, and $h$ is the step size used in the spatial discretization.

Also for $\rho_1$, the scheme used for the time evolution is of second order in the numerical time step.
The evolution dictated by the first two terms on the right hand side of Eq.~(\ref{TDLE_1part}) may be implemented by means analogous to the one-particle part of the propagator for the two-particle pure state. The $\rho_1$-propagator also features terms originating directly from the source term and cross-terms between the source term and the effective Hamiltonian.
See App.~\ref{App_Propagator} for more detail.

We have also calculated the time-dependent zero-particle probability $p_0$, which is obtained from Eq.~(\ref{TDLE_nPart}) with $n=0$. This amounts to simply time-integrating the source term
\begin{equation}
\label{VacuumSource}
\mathcal{S}[\rho_1(t)] = 2 \int \gamma(x) \rho_1(x, x) \, dx , 
\end{equation}
which is a scalar. While this does not provide any differential information about any absorbed particles, it serves as a useful check for the numerics.
The Lindblad equation ensures that the trace of the total density matrix remains unity. In this context, this means that
\begin{equation}
\label{ConserveTrace}
|\Psi_2(t)|^2 + \mathrm{Tr} \rho_1(t) + p_0(t) = 1
\end{equation}
at all times.
As our numerical scheme is not manifestly trace conserving,
the deviation from unity 
is a measure of the accuracy of
simulations.

In our examples, we set out to calculate the energy distribution of the unbound particles emerging after interaction.
We do so by projection onto the eigen-states of the unperturbed, Hermitian one-particle Hamiltonian $h_0$ as dictated by Eqs.~(\ref{dPdE_General}, \ref{dPdE_GeneralOne}).
With our matrix representation of the two-particle wave function,
the spatial integral in Eq.~(\ref{PhiDef}) may
be found by matrix multiplication.
Numerically, the effective one-particle density matrix of Eq.~(\ref{PhiDef}) may be calculated as a sum of matrix products,
\begin{equation}
\label{PhiAsMatrix}
\Phi = h \tau \, \sum_{t_n} \, \Psi_2(t_n) \Psi_2^\dagger(t_n),
\end{equation}
where $\tau$ is the temporal step size.
Eq.~(\ref{dPdE_General}) may conveniently be expressed as
\begin{equation}
\label{dPdE_Matrix}
\hbar \frac{\partial P_2}{\partial \varepsilon} \approx 2 h^2 \, \boldsymbol{\varphi}_\varepsilon^\dagger (D \Phi + \Phi D) \boldsymbol{\varphi}_\varepsilon ,
\end{equation}
where $\boldsymbol{\varphi}_\varepsilon$ represents the scattering state according to the eigen energy $\varepsilon$ as a column vector.
The $\varepsilon$-differential distribution obtained from $\rho_1$ is calculated analogously to Eq.~(\ref{dPdE_Matrix}) -- with
$2 \Phi$ replaced by the time-integral of $\rho_1$.
Apart from certain increase in memory requirements,
calculating the effective one-particle density matrix $\Phi$ when solving Eq.~(\ref{TDSE}) and the time-integral of $\rho_1$ when solving Eq.~(\ref{TDLE_1part}) impose little extra numerical effort.
Nor does acquiring the time-independent one-particle scattering states $\boldsymbol{\varphi}_\varepsilon$ impose any substantial workload.

When using numerical box-normalized eigen states of $h_0$ instead of true continuum states in interpolating the continuous distribution of Eq.~(\ref{dPdE_Matrix}),
we must ensure correct normalization. This means that the projections must be multiplied by the {\it density of states} before interpolation. Moreover, we distinguish between the two channels consisting of symmetric and anti-symmetric scattering states and add their respective contributions to the total spectrum incoherently.

\subsection{Convergence of the energy spectra}
\label{Sec_Convergence}

\begin{figure}
  \centering
  \includegraphics[width=6cm]{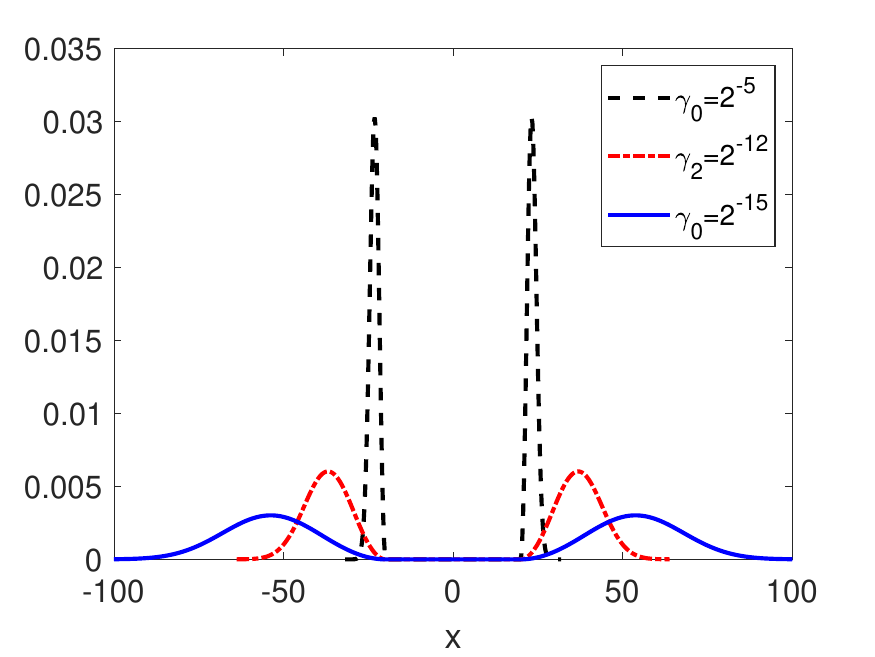}
  \caption{The outgoing waves which are accumulated in time and analyzed according to Eqs.~(\ref{dPdE_General}) and (\ref{dPdE_GeneralOne}) constitute effective one-particle matrices. The plot shows the diagonal of the matrix which enters into Eq.~(\ref{dPdE_General}), $\mathrm{diag} \left[ \{\gamma(x) + \gamma(x')\} \Phi(x,x') \right]$, for three different choices of the CAP strength $\gamma_0$, cf. Eqs.~(\ref{PhiDef}) and (\ref{CAPfunction}). The CAP onset $x_0=20$. With a strong CAP, the outgoing waves do not reach very far into the CAP region before being attenuated which, in turn, causes a narrow confinement in space. This particular two-particle system correspond to that of Sec.~\ref{Sec_PhotoIonExample}.}
  \label{Fig_WidthOfAbs}
\end{figure}

In a numerical simulation it is desirable to maintain as hard absorption as possible as this allows for a strongly truncated numerical domain. On the other hand, the absorption must also be sufficiently soft to ensure that the results do not depend on the characteristics of the CAP itself. In our numerical examples we check for convergence in the absorber strength; $\gamma_0$ in Eq.~(\ref{CAPfunction}) is extrapolated towards zero. We run a rather large number of simulations with decreasing $\gamma_0$-values in order to investigate the transition from CAP-dependent spectra towards CAP independent ones in some detail. However, some remarks about the admissible magnitude of $\gamma_0$ may also be made {\it a priori}. 
For instance, the absorption cannot be so strong that it induces reflections. Moreover, too hard absorption leads to poorly resolved energy spectra
due to the Heissenberg principle.
If the accumulated absorbed waves which enter into Eqs.~(\ref{dPdE_General}, \ref{dPdE_GeneralOne}) have very
narrow spatial confinements, the corresponding calculated energy distributions may be unable to resolve the true energy spectra.
Suppose that the physical nature of the process under study is such that energy spectra require a resolution given by $\Delta \varepsilon$ and that the accumulated outgoing waves has the width $\Delta x$ -- in each direction.
Then we must require that the spatial extension fulfills
\begin{equation}
\label{HeissenbergRestrictionBox}
\Delta x \gg \hbar \sqrt{\frac{\varepsilon}{2 m}} \, \frac{1}{\Delta \varepsilon} .
\end{equation}
As illustrated in Fig.~\ref{Fig_WidthOfAbs}, the extension $\Delta x$ is directly determined by the CAP strength. Correspondingly, the inequality~(\ref{HeissenbergRestrictionBox}) effectively imposes an upper bound on $\gamma_0$ in Eq.~(\ref{CAPfunction}) for the specific system we wish to describe.
It should also be noted that the physical nature of this system may add further restrictions on the CAP strength.

\subsection{Example I: Scattering}
\label{Sec_ScatteringExample}

\begin{figure}
  \centering
  \includegraphics[width=6cm]{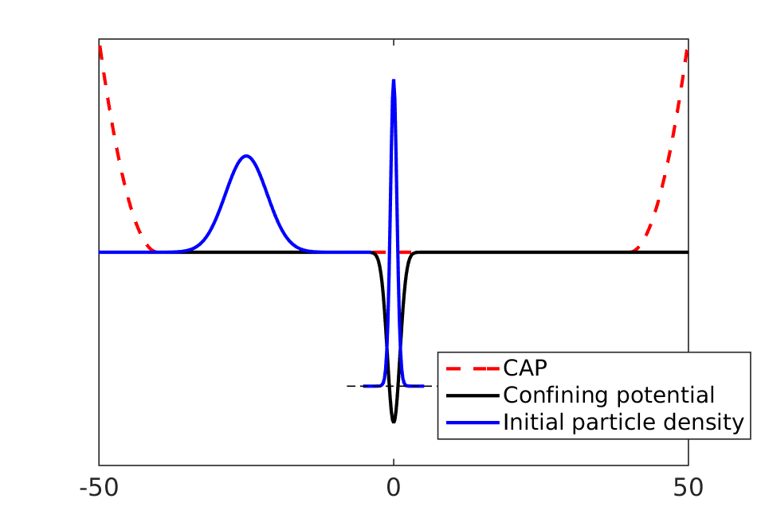}
  \caption{The figure illustrates the situation prior to collision. A projectile particle of rather sharply defined initial momentum is incident upon a target particle initially trapped in the ground state of a confining potential.}
  \label{Fig_ScatteringSetup}
\end{figure}

In this example a target particle is initially trapped in the ground state of a short-ranged Gaussian potential,
\begin{equation}
\label{ConfiningPot}
V(x) = -V_0 \exp\left( -\frac{x^2}{2 \sigma_V^2}\right) ,
\end{equation}
while an identical projectile particle with a Gaussian wave packet is incident on the target,
see Fig.~\ref{Fig_ScatteringSetup}.
The situation could correspond to a quantum dot embedded in a quantum wire \cite{Gumbs1999,Bednarek2003,Ciftja2006,Selsto2013,Pont2016}.

We take our units to be defined by setting $\hbar$ and the particle mass to unity.
The confining potential has the strength $V_0=4$ and the width $\sigma_V=3/(2\sqrt{2})$ in these units, cf. Eq.~(\ref{ConfiningPot}). This leads to a one-particle ground state energy of $-3.141$. The interaction strength $W_0=1$, and for the softening parameter $s$, the value 0.1925 has been used, cf. Eq.~(\ref{Interaction}). The square CAP function is nonzero for $|x|$ beyond $x_0=35$, cf. Eq.~(\ref{CAPfunction}). The initial Gaussian projectile wave function is centred at $x=-20$ in position space and $p=2$ in momentum space. Its momentum width is 0.1, which corresponds to a position width of 5 length units.

Since the initial energy of the projectile is such that it hardly allows for both particles to be liberated, it suffices to solve Eq.~(\ref{TDSE}); we do not need to consider any second absorption.

\begin{figure}
  \centering
  \begin{tabular}{c}
  \includegraphics[width=6cm]{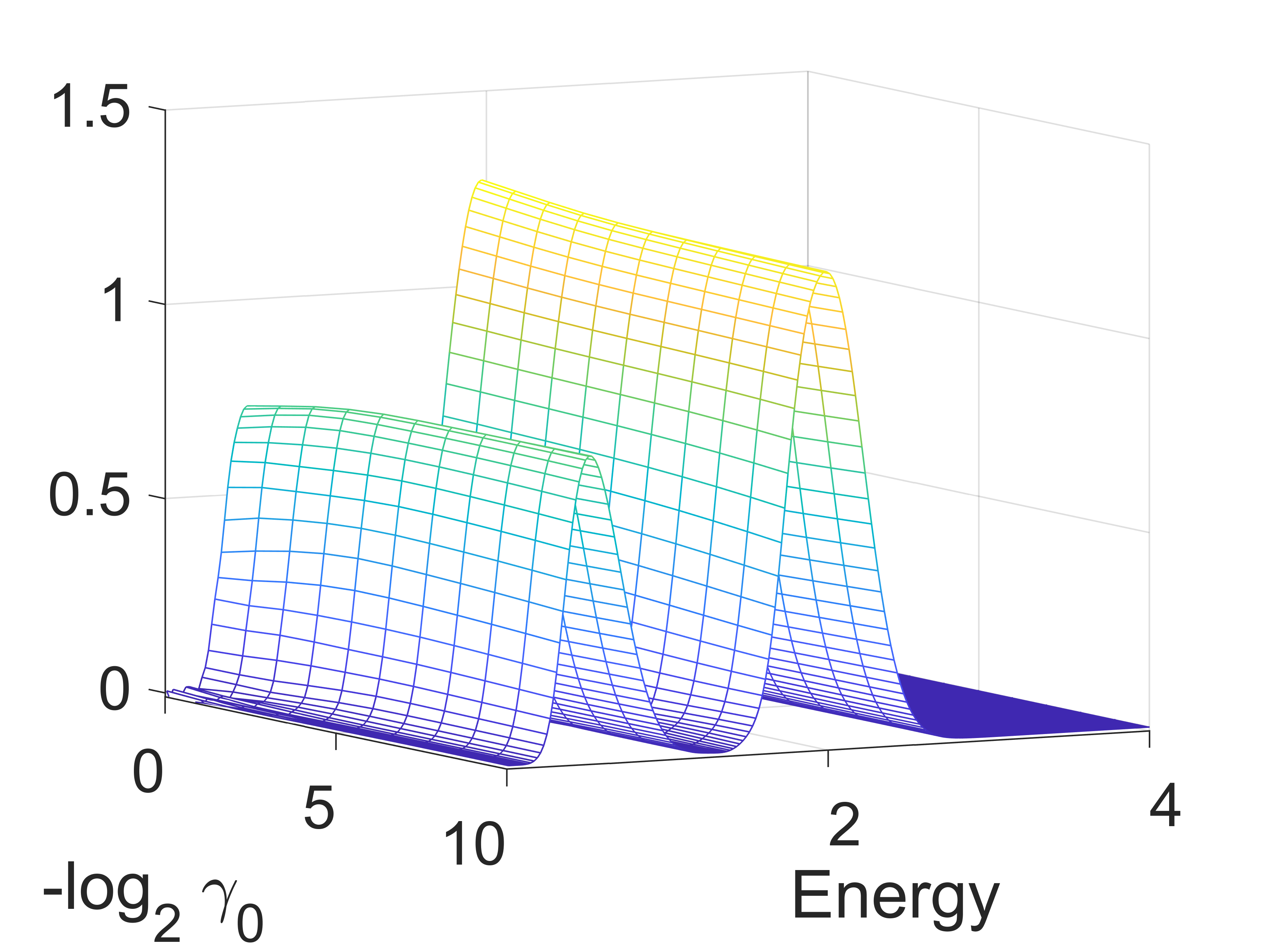} \\
  \includegraphics[width=6cm]{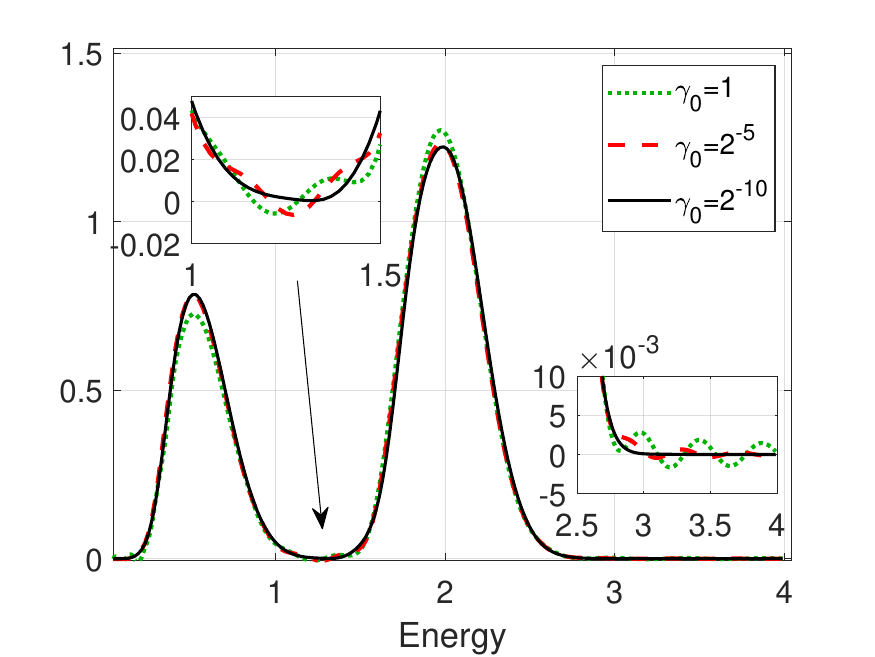} \\
   \includegraphics[width=6cm]{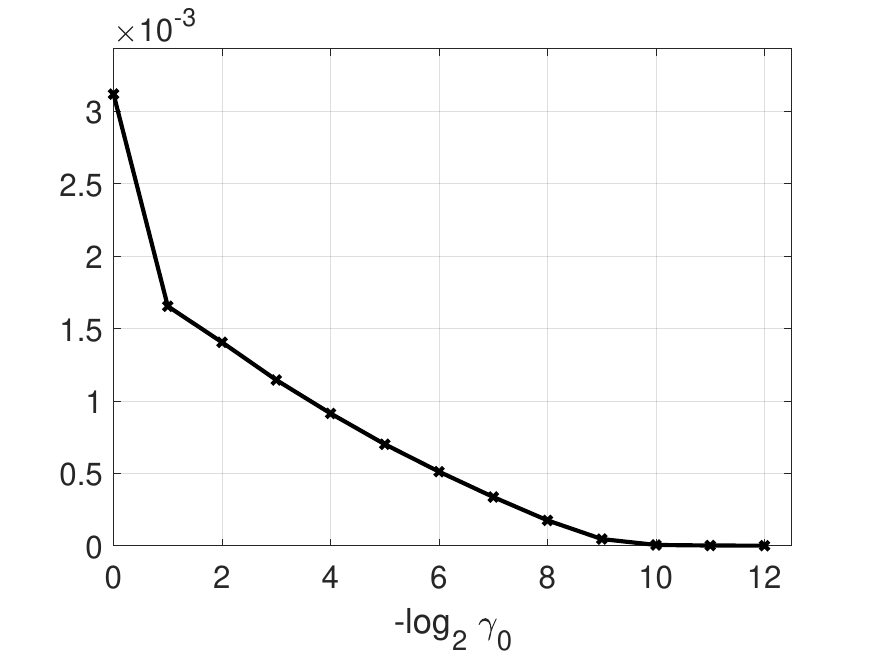} \\
  \end{tabular}
  \caption{{\it Upper panel}: The energy spectrum of the unbound particle emerging from the collision event. In this case, the projectile particle has an initial mean momentum of 2 units and the target particle is trapped in the ground state with energy $-3.14$ units. The spectrum is predicted by Eq.~(\ref{dPdE_General}) and calculated for various values of the strength of the CAP function, i.e., $\gamma_0$ in Eq.~(\ref{CAPfunction}).
  {\it Middle panel}: The same distribution as in the upper panel for three values of $\gamma_0$. The inserts display close-ups on regions in which unconverged spectra feature negative values.
  {\it Lower panel}: In this plot, the total negative contribution to the (unconverged) energy distributions is plotted as a function of absorber strength.}
  \label{Fig_ScatteringExample}
\end{figure}

The upper panel of Fig.~\ref{Fig_ScatteringExample} shows the energy distribution of the particle emerging from the collision event as predicted by Eq.~(\ref{dPdE_General}). In addition to energy, it is also shown as a function of the CAP strength $\gamma_0$, cf. Eq.~(\ref{CAPfunction}). The values of this parameter are chosen
such that $\gamma_0=2^{-n}$ where $n$ is a non-negative integer. We see that not only does the distribution converge as $\gamma_0$ decreases; it is in fact virtually independent of $\gamma_0$. Only for very strong absorption can we see deviations from the converged one.

In the middle panel of Fig.~\ref{Fig_ScatteringExample}, the same energy spectra are shown for three values of $\gamma_0$. We see that the wave emerging from the collision event comes ut in two energy lobes. This is due to the fact that the target has, with a certain probability, been excited. In other words, the peak centred around 2 energy units corresponds to elastic scattering while the peak near 0.5 energy units corresponds to inelastic scattering.

The inserts in the panel reveal that the spectra obtained with comparatively hard absorption are not strictly non-negative; in certain regions they are negative.
This is related to the fact that, as discussed in Sec.~\ref{Sec_AbsFromN}, the operator $\{ \Gamma, \rho_N \}$ in Eqs.~(\ref{ExpectationValueRemoved}) and (\ref{ExpectationValueRemovedSinglePart}) is not necessarily positive semi-definite.
Thus, there is no obvious mathematical reason why differential quantities obtained from Eq.~(\ref{dPdE_General})
must be non-negative.
However, in the limit $\gamma_0 \rightarrow 0^+$, when the predictions become physical, no negative parts are seen. This is demonstrated more explicitly in the
lower panel of Fig.~\ref{Fig_ScatteringExample}, which depicts the unphysical negative contribution to the energy distribution $\partial P_2/\partial \varepsilon$,
\begin{equation}
\label{NegativeContributionInt}
-\int_{\partial P_2/\partial \varepsilon <0} \frac{\partial P_2}{\partial \varepsilon} \, d \varepsilon ,
\end{equation}
as a function of $\gamma_0$.
We clearly see that this undesirable feature in fact vanishes for finite values of $\gamma_0$.

\begin{figure}
  \centering
  \begin{tabular}{lc}
  \includegraphics[width=4.2 cm]{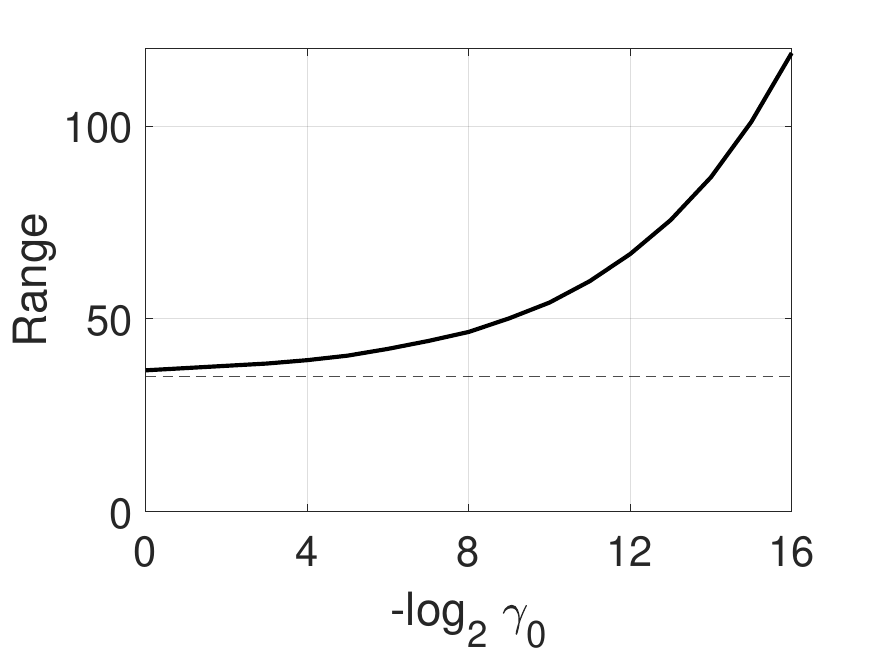} &
  \includegraphics[width=4.2 cm]{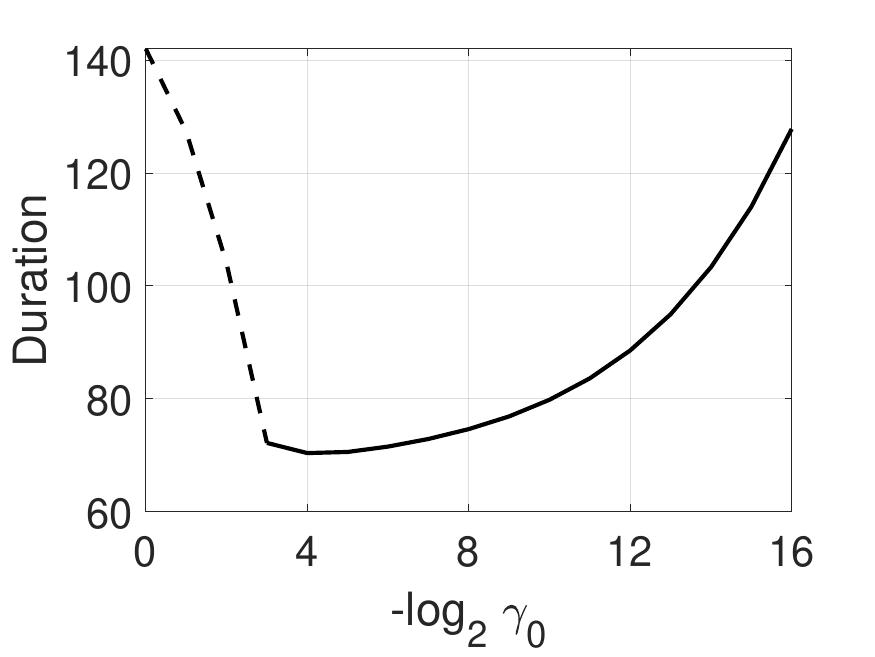}
  \end{tabular}
  \caption{These panels show the maximum range of the wave function ({\it left}) and the duration of the simulation ({\it right}) for the collision example pertaining to Fig.~\ref{Fig_ScatteringExample}. For the sake of illustration we have included calculations both featuring much weaker absorption than necessary and unreasonably hard absorption. The horizontal dashed line in the left panel indicates the onset of the CAP function, and the dashed part of the curve in the right panel indicates the region in which artificial reflection effectively prolongs the simulations.}
  \label{Fig_RangeAndDuration}
\end{figure}

As mentioned, it is desirable to use as hard absorption as possible -- while still obtaining converged spectra and avoiding artificial reflections. In Fig.~\ref{Fig_RangeAndDuration} we display how the extension of the wave function, which is subject to absorption, and the duration of the simulation depends on the strength of the absorber.
We have here defined the former as the smallest $a$ which is such that whatever resides beyond
$|x| = a$ has a squared norm less than 1 \% throughout the simulation.
The duration we have defined as the time it takes for $|\Psi_2(t)|^2$ to fall below 1~\%. The left panel shows, as expected, that a larger grid is necessary as the CAP strength is reduced. When it comes to the duration of the simulation, however, the behaviour is not monotonous; initially, the time it takes to simulate the entire event actually decreases with decreasing $\gamma_0$. This is due to artificial reflections induced by too hard absorption. Reflected waves will have to travel back across the grid at least one more time before being absorbed, and, thus, complete absorption takes longer in this case.

\begin{figure}
  \centering
  \begin{tabular}{lc}
  \includegraphics[width=4.2 cm]{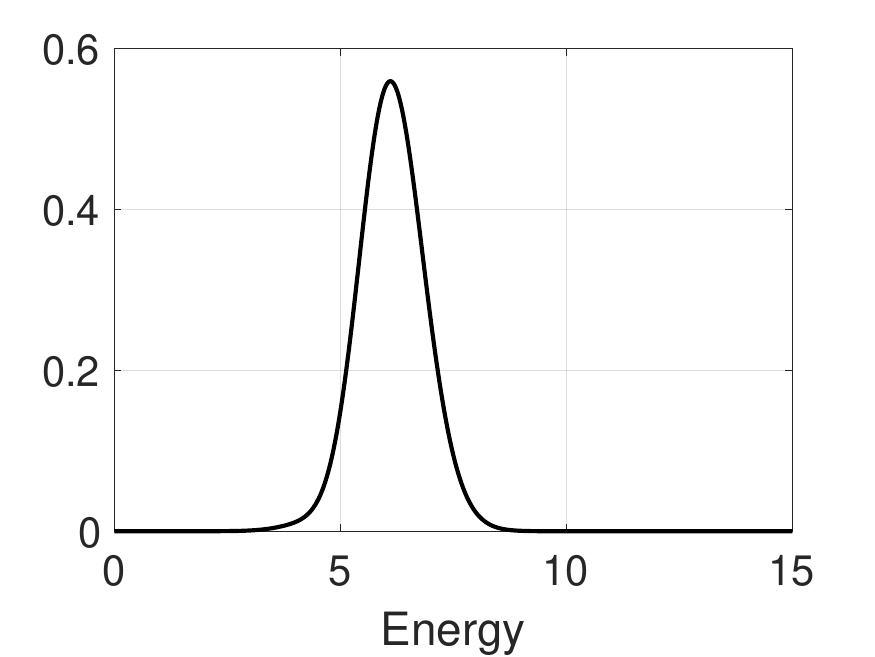} &
  \includegraphics[width=4.2 cm]{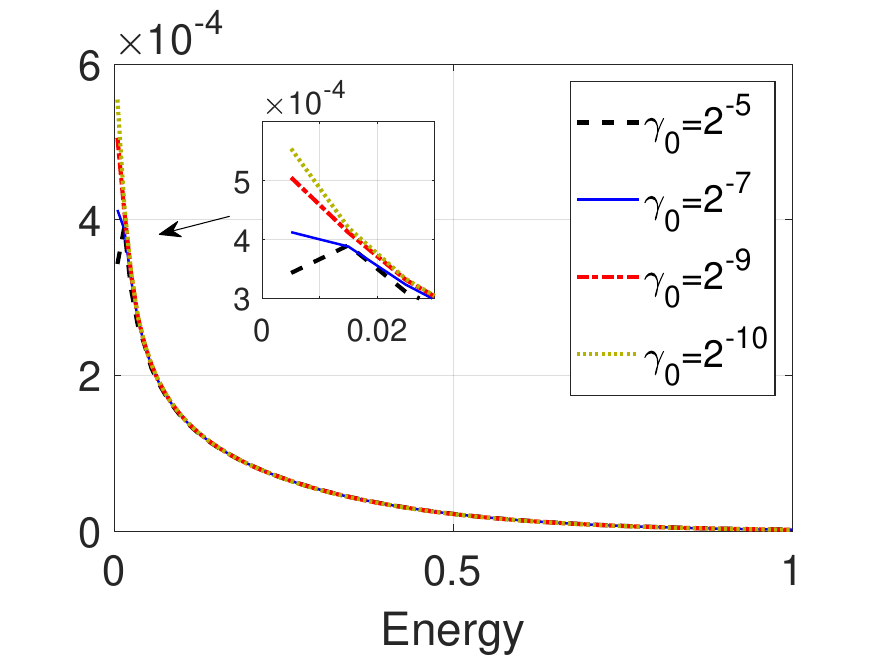}
  \end{tabular}
  \caption{The left panel, which is analogous to the middle panel of Fig.~\ref{Fig_ScatteringExample}, shows the energy distribution of the first particle to be absorbed after collision. Here the mean momentum of the projectile particle is 3.5 units, and its width is 0.2. The right panel displays the energy distribution of the second particle to be absorbed. It is not likely that two liberated particles will emerge from the collision; this distribution integrates to a total probability of $5.4 \cdot 10^{-5}$. The distribution in the right panel features a certain dependence of the strength of the CAP function, $\gamma_0$ in Eq.~(\ref{CAPfunction}), near threshold. Otherwise, it is virtually CAP independent.
  }
  \label{Fig_ScatteringExampleV2}
\end{figure}

In Fig.~\ref{Fig_ScatteringExampleV2} we display the results of a collision event for which the liberation of both particles is energetically admissible. In this case the projectile particle has an initial momentum with a mean value of 3.5 units and a width of 0.2 units. The left panel of Fig.~\ref{Fig_ScatteringExampleV2} reveals that elastic scattering is the dominant process. However, as the right panel shows, there is a certain probability for the second particle to be liberated as well. The corresponding energy distribution is obtained from Eq.~(\ref{dPdE_GeneralOne}), which, in turn, requires the solution of Eq.~(\ref{TDLE_1part}) in addition to Eq.~(\ref{TDSE}).
It is seen that this second particle predominantly comes out with low energy.
It is also seen that also this energy distribution is very weakly dependent on $\gamma_0$.
We do see some dependence, however, close to threshold.
This is related to the fact that particles with near-zero energy require a very long time to reach the absorber; complete absorption is hard to achieve in a simulation of finite duration in this case.
It may seem counter-intuitive that this issue is more prominent at harder absorption than softer absorption. It can be understood from what we saw in the right panel of Fig.~\ref{Fig_RangeAndDuration}. Low-energy waves are more prone be reflected by the CAP than faster ones -- and increasingly so with harder absorption. Consequently, with a finite duration of simulations, 1000 time units in this case, the slow, reflected waves do not have enough time hit the absorber many enough times to reach full absorption.

\subsection{Example II: Photo ionization}
\label{Sec_PhotoIonExample}

The next example addresses a model for photo ionization of a two-electron atom. The electrons are initially confined in the two-particle ground state of a regularized Coulomb potential with a form identical to the interaction Eq.~(\ref{Interaction}):
\begin{equation}
\label{ConfiningPotCoulomb}
V(x) = - \frac{V_0}{\sqrt{x^2+u^2}} .
\end{equation}
Numerically, this initial state is constructed by evolving the system without the CAP in {\it imaginary time}, i.e., by substituting $t$ with $-i t$ in Eq.~(\ref{TDSE}) with $\Gamma=0$. This, along with renormalization at each time step, causes virtually any initial state to converge towards the ground state exponentially.

Next, the system is exposed to a laser pulse. The interaction with the laser is formulated in the velocity gauge, i.e., the Hermitian part of the one-particle hamiltonian reads
\begin{equation}
\label{OnePartHam}
h^{(i)} = \frac{p_i^2}{2m} +V(x_i) + \frac{e}{m} A(t) p_i,
\end{equation}
where
$i=1,2$ refers to the particle number, $m$ is the electron's mass and $-e$ is its charge.
The homogeneous vector potential $A$ reads
\begin{equation}
\label{LaserPulse}
A(t) = \left\{
\begin{array}{lc}
\frac{E_0}{\omega} \sin^2 \left( \frac{\pi}{T}\omega t \right) \sin \left( \omega t \right) , & 0 \leq t \leq T\\
0, & \text{otherwise}
\end{array}
\right. .
\end{equation}

In this particular example we apply atomic units, a.u., which may be defined by, in addition to setting $\hbar$ and $m$ to unity as in Sec.~\ref{Sec_ScatteringExample}, choosing the elementary charge $e$ and the Bohr radius as units for their respective quantities.
Here, the peak electric field strength is $E_0=0.1$~a.u., the central angular frequency $\omega$ is $0.3$~a.u., and the duration $T$ corresponds to seven optical cycles. The confining potential, Eq.~(\ref{ConfiningPotCoulomb}), is chosen such that both $V_0$ and $u$ are 0.5~a.u..
This yields a one-particle ground state energy of $-1/2$~a.u.
Also for the interaction, Eq.~(\ref{Interaction}), we have set the parameters $W_0$ and $s$ to $0.5$~a.u.. The resulting two-particle ground state energy is $-0.554$~a.u.. Thus, one particle is rather weakly bound.
The CAP is turned on at $|x|=x_0=50$ atomic length units.

\begin{figure}
  \centering
  \begin{tabular}{lr}
  \includegraphics[width=4.2cm]{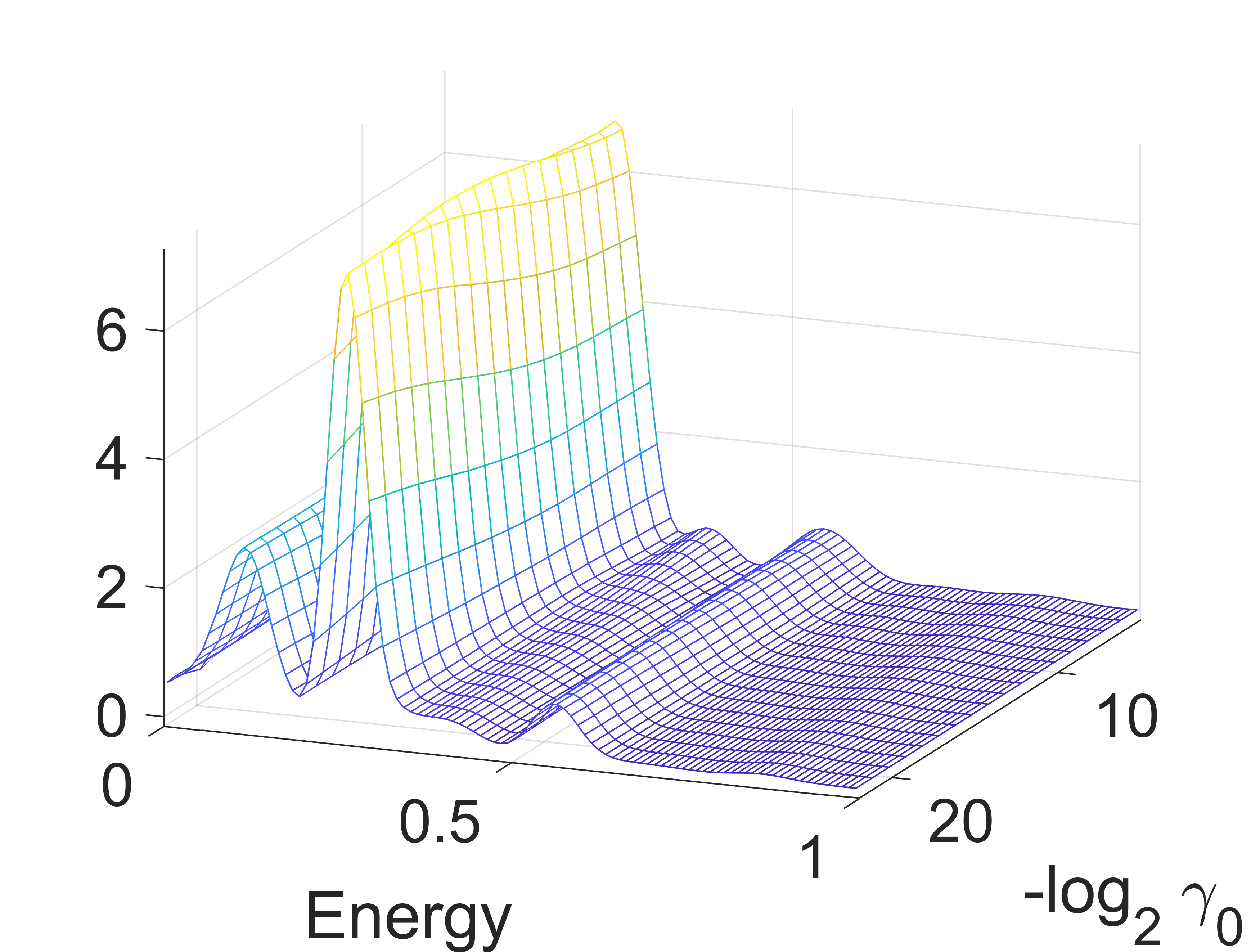} &
  \includegraphics[width=4.2cm]{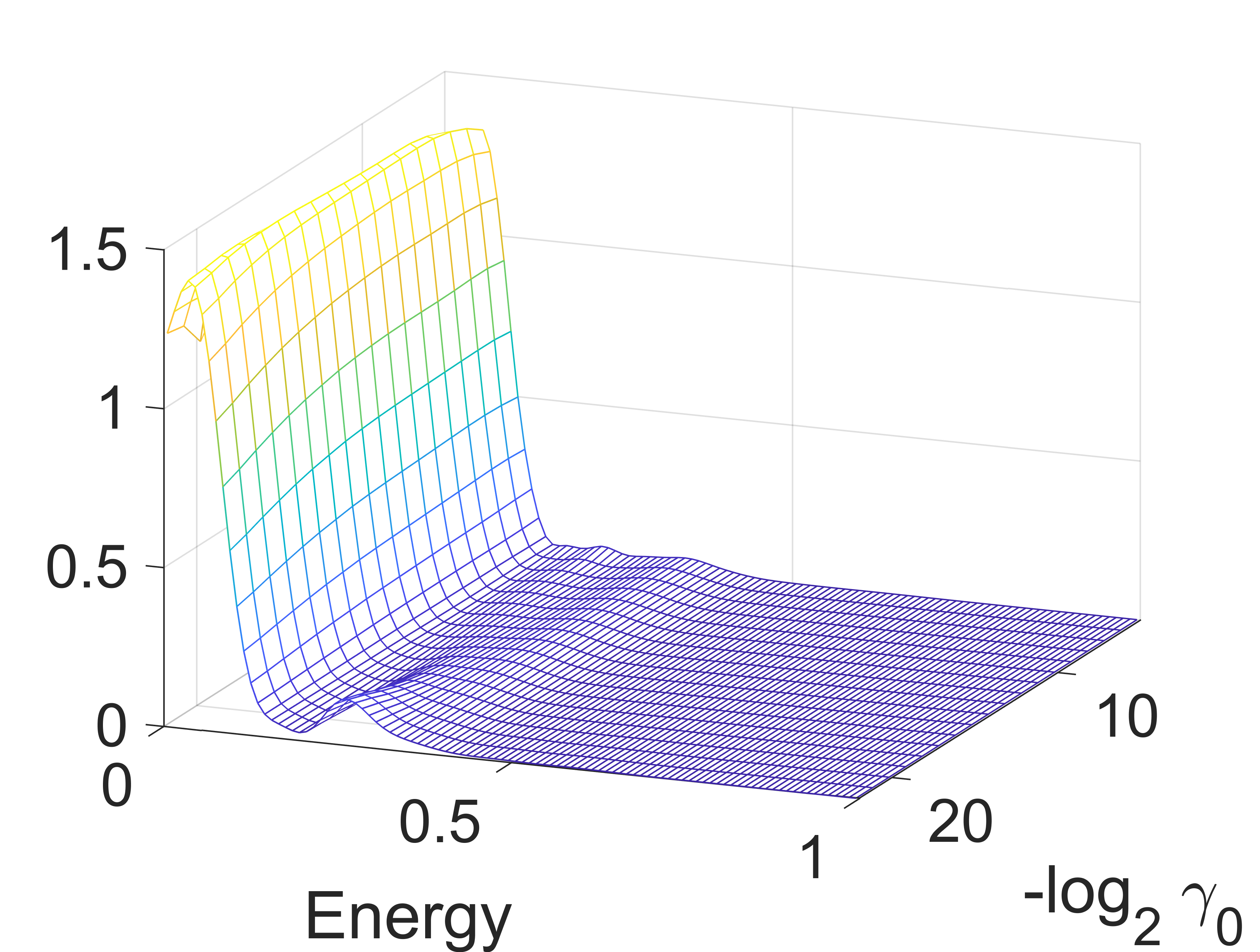}
  \end{tabular}
  \caption{These plots show the photo electron spectra for a model two-electron atom exposed to a laser pulse. The energy required for single ionization is $0.054$~a.u., double ionization requires $0.554$~a.u. and the laser pulse corresponds to a photon energy of $0.3$~a.u., see text for more details on the system. The spectra are calculated with various absorber strengths, i.e., $\gamma_0$ in Eq.~(\ref{CAPfunction}). The left panel is obtained from absorption from the original two-particle system, while the rights panel is obtained from the one-particle sub-system which remains after the first absorption event.}
  \label{Fig_ConvergencePhotoEl}
\end{figure}

Figure~\ref{Fig_ConvergencePhotoEl} demonstrates, analogously to the upper panel of Fig.~\ref{Fig_ScatteringExample}, how the predicted singly-differential photo-electron spectra depend on the strength of the CAP function. The left panel is the spectrum obtained from the first absorption, calculated using Eq.~(\ref{dPdE_General}), and the right one corresponds to the second absorption, calculated using Eq.~(\ref{dPdE_GeneralOne}).
The
spectrum obtained from the first absorption is not as close to being $\gamma_0$-independent as the one in Fig.~(\ref{Fig_ScatteringExample}). We can, e.g., detect a slight shift towards higher energies as $\gamma_0$ approaches zero. However, the dependence on the CAP strength is still quite weak, and the spectrum does converge rather rapidly as the CAP strength diminishes.

\begin{figure}
  \centering
  \includegraphics[width=6cm]{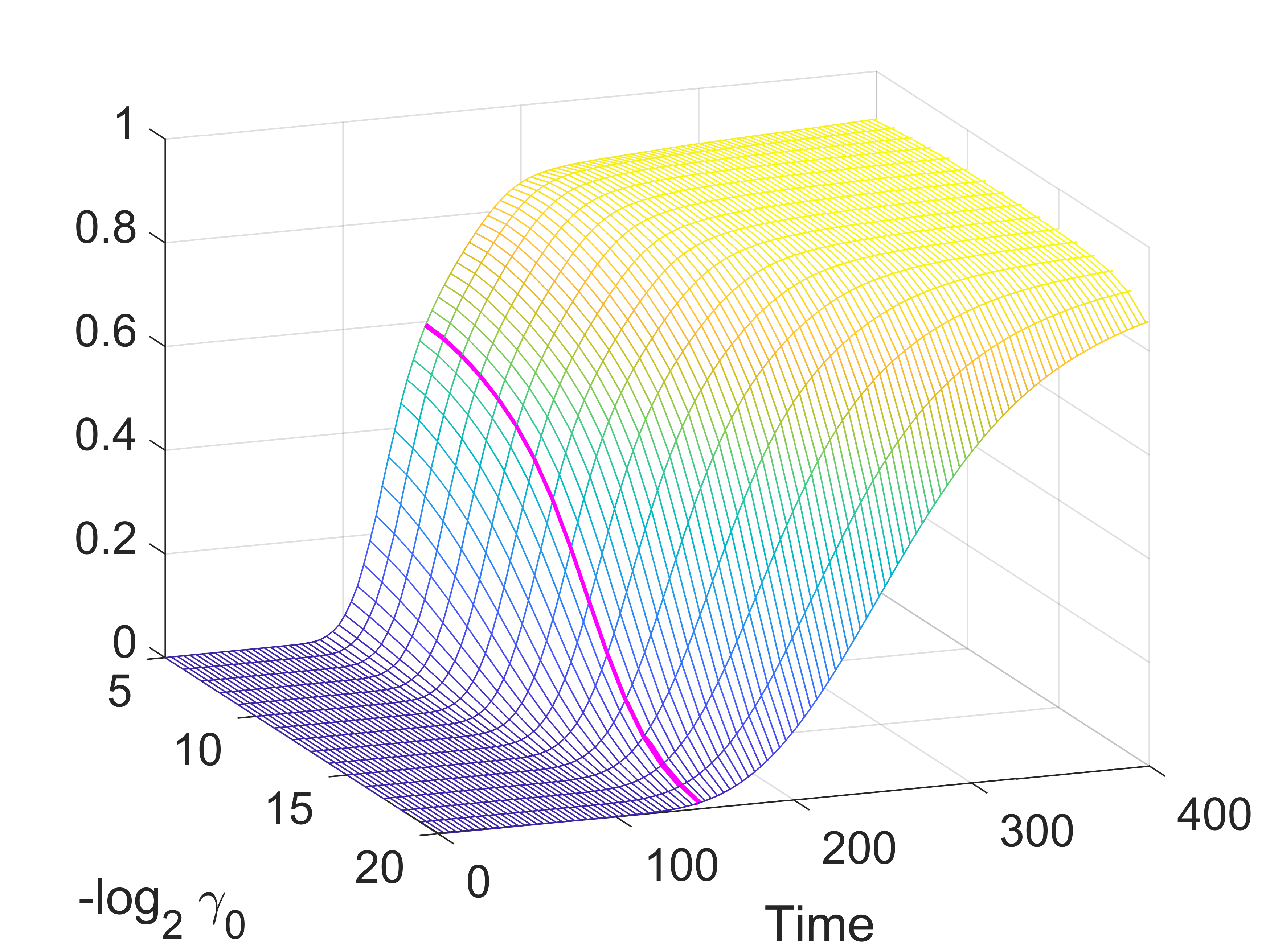}
  \caption{The total absorption from the two-particle system, $1 - |\Psi_2(t)|^2$, for the system of Fig.~\ref{Fig_ConvergencePhotoEl} as a function of time and absorber strength $\gamma_0$. The thick, purple curve indicates the time at which the laser pulse is over.}
  \label{Fig_AbsorbTimeGamma0}
\end{figure}

Interestingly, this happens despite the fact that much of the absorption takes place {\it during} interaction with the laser pulse. This is illustrated in Fig.~\ref{Fig_AbsorbTimeGamma0}, which depicts the depletion in norm from $\Psi_2$ as a function of time, i.e., it shows $1-|\Psi_2(t)|^2$ as a function of $t$ and $\gamma_0$. The thick purple curve corresponds to the time at which the pulse is switched off.
By comparing Fig.~\ref{Fig_AbsorbTimeGamma0} with the left panel of Fig.~\ref{Fig_ConvergencePhotoEl} it is seen that a converged spectrum is obtained before the laser interaction is over. This may seem odd for various reasons. Due to the explicit time-dependence in the Hamiltonian, it would, e.g., seem more reasonable to apply Eq.~(\ref{dPdE_Time}), which involves time-dependent scattering states, rather than Eq.~(\ref{dPdE_General}). And even doing so, by absorbing an electron you would still deprive it of the possibility to exchange energy with the laser field.
The latter suggests that such exchange predominantly takes place within the CAP free region -- close to the centre of the Coulomb-like potential.
The fact that Eq.~(\ref{dPdE_General}), in which the absorbed wave is projected onto time-independent scattering states, indeed produces a converged spectrum 
despite absorption 
during explicitly time-dependent interaction is related to the fact that the interaction is described in the velocity gauge. In this formulation, a free, classical electron is at rest in momentum space; the momentum is a constant of motion. This is reflected in the fact that the free-electron Volkov solutions are time-independent -- apart from a phase factor which does not contribute in this density-matrix formalism~\cite{Tao2012}. In fact, if we substitute our scattering states $\varphi_{\varepsilon}$ with solutions in which the confining potential is removed, i.e., plane waves, we would obtain a similar spectrum -- except for a shift towards higher energies for the multi-photon peaks and some irregularities at the low-energy part of the spectrum. Both of these deficiencies are due to the neglect of the Coulomb-like potential, and both diminish as $\gamma_0$ decreases since this causes absorption to take place further away from the centre of the Coulomb-like potential.

While the energy distribution of the first electron to be absorbed, $\partial P_2/ \partial \varepsilon$, converges rather quickly as the CAP strength is reduced, this 
convergence is somewhat slower 
for the second particle to be absorbed. The right panel of Fig.~\ref{Fig_ConvergencePhotoEl} shows $\partial P_1 / \partial \varepsilon$ obtained from Eq.~(\ref{dPdE_GeneralOne}). Although the first peak near threshold is rather well resolved with comparatively strong absorption, the peak centred near $\varepsilon = 0.25$~a.u., 
requires a weaker CAP strength for 
convergence.

\begin{figure}
  \centering
  \begin{tabular}{lr}
  \includegraphics[width=4.2cm]{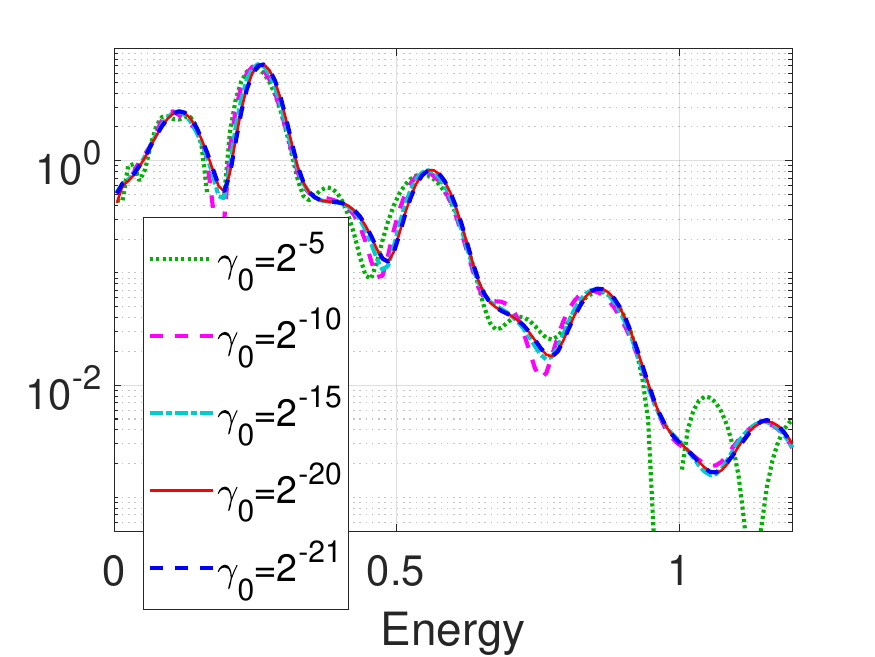} &
  \includegraphics[width=4.2cm]{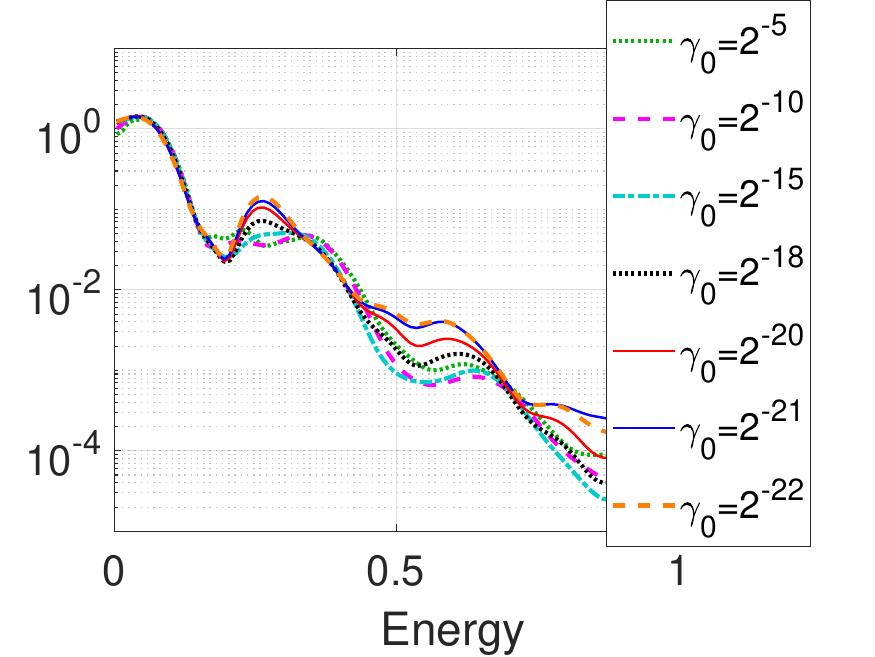}
  \end{tabular}
  \caption{Here the same photo electron spectra as in Fig.~\ref{Fig_ConvergencePhotoEl} are shown for certain values of the CAP strength $\gamma_0$, however using a logarithmic $y$-axis. As in Fig.~\ref{Fig_ConvergencePhotoEl}, the left panel stems from the first absorption while the right one is calculated from the second absorption. The green curve, which corresponds to rather strong absorption, is seen to vanish in certain regions. This is because too strong absorption may induce negative values for the spectrum, cf. the middle and lower panel of Fig.~\ref{Fig_ScatteringExample}.}
  \label{Fig_ConvergencePhotoElLogScale}
\end{figure}

Fig.~\ref{Fig_ConvergencePhotoElLogScale}, which shows the same spectra as in Fig.~\ref{Fig_ConvergencePhotoEl} with a logarithmic $y$-axis for certain values of $\gamma_0$, shows more details in this regard.
The spectrum of the first absorbed particle, $\partial P_2/\partial \varepsilon$, features several well converged multi-photon peaks. 
Most of these peaks correspond to a single-ionization process in which the remaining ion is left in its ground state. 
spectrum obtained from the second absorbtion, $\partial P_1/\partial \varepsilon$, 
does not feature equally pronounced multi-photon peaks. The maximum just above threshold is consistent with a direct two-photon double-ionization process. 
Moreover, it is also interesting to note that the $\gamma_0$-dependence in $\partial P_1 / \partial \varepsilon$ seems to be more prominent at higher energies.

\begin{figure}
  \centering
  \begin{tabular}{lr}
  \includegraphics[width=4.2cm]{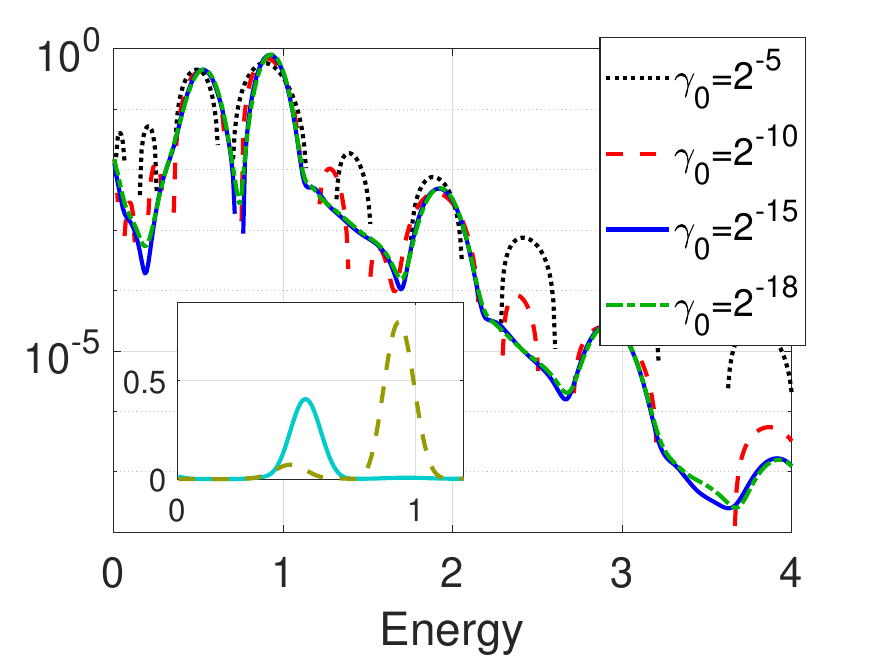} &
  \includegraphics[width=4.2cm]{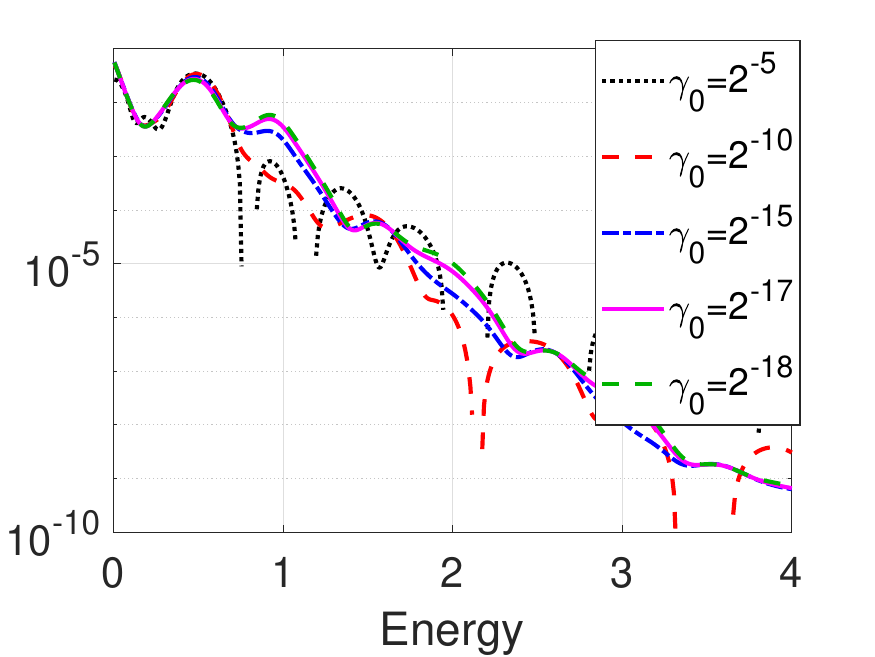}
  \end{tabular}
  \caption{The same kind of photo electron spectra as in Fig.~\ref{Fig_ConvergencePhotoElLogScale} -- however with a maximum electric field strength of $E_0=0.25$~a.u., a central frequency of $\omega=1$~a.u. and a duration corresponding to 10 optical cycles in this case, cf. Eq.~(\ref{LaserPulse}). Here, the CAP function sets on at $x_0=20$~a.u.. The left panel corresponds to the predictions of Eq.~(\ref{dPdE_General}), while the right panel is obtained using Eq.~(\ref{dPdE_GeneralOne}). The insert in the left panel shows partial contributions from symmetric (full curve) and anti-symmetric (dashed curve) scattering states. In the insert linear axes are used.}
  \label{Fig_ConvergencePhotoOmegaOneLogScale}
\end{figure}

Figure~\ref{Fig_ConvergencePhotoOmegaOneLogScale} is analogous to Fig.~\ref{Fig_ConvergencePhotoElLogScale}. It shows photoelectron spectra for the same system, however, this time it is exposed to a laser pulse with the central frequency $\omega=1$~a.u., the peak electric field strength $E_0=0.25$~a.u. and a duration corresponding to 10 optical cycles, $T= 10 \times 2 \pi/\omega$, cf. Eq.~(\ref{LaserPulse}). The corresponding photon energy, $\hbar \omega$, 
allows for one-photon double ionization, as opposed to the previous case.
Here, we have set the onset of the CAP region 
at $x_0=20$~a.u..
From the first absorption, we again see pronounced single-ionization peaks. 
And, again, most of these correspond to single ionization with the remaining ion in the ground state. From energy-considerations, the peak at about 0.5~a.u., however, seems to predominantly be due to single ionization combined with excitation of the remaining electron to its first excited state. This is consistent with calculated population of the first excited state within the one-particle sub-system,
\begin{equation}
\label{ProbFirsExcited}
\langle \varphi_\mathrm{1st} | \rho_1(t \rightarrow \infty) | \varphi_\mathrm{1st} \rangle
= 0.060 ,
\end{equation}
where $\varphi_\mathrm{1st}$ is the wave function of the first excited one-particle state.
It is also consistent with the fact that the peak at $\varepsilon \approx 0.5$~a.u. is dominated by symmetric scattering states -- as opposed to anti-symmetric ones in the case of single ionization with relaxation to the (symmetric) ground state. This is displayed in the insert in the left panel of Fig.~\ref{Fig_ConvergencePhotoOmegaOneLogScale}.

Although not equally pronounced as for the single-ionization case, multi-photon peaks may also be seen in the spectrum obtained from the second absorption, $\partial P_1 / \partial \varepsilon$. The peaks seen at $n \times \hbar \omega - 0.5~\text{a.u.}$ correspond to $n$-photon ionization from the ground state of the ion remaining after the first ionization, i.e., these are sequential double ionization processes. The peak at $\varepsilon \approx 0.90$~a.u. is consistent with a sequential double ionization process via the first excited state of the ion. As in the case with $\omega=0.3$~a.u., we see significant contributions at very low energies.

When it comes to $\gamma_0$-dependence, Fig.~\ref{Fig_ConvergencePhotoOmegaOneLogScale} shows the same tendency as Fig.~\ref{Fig_ConvergencePhotoElLogScale} in that the spectrum $\partial P_1 / \partial \varepsilon$ requires a weaker CAP in order to become $\gamma_0$-independent than does $\partial P_2 / \partial \varepsilon$. This observation is 
consistent with the bias introduced in the sequential double absorption process: The fastest electron reaches the absorber first. In situations such as these ones, where both electrons of $\Psi_2$ may have reached their respective continua simultaneously, the most energetic of the two will predominantly contribute to the $\partial P_2/\partial \varepsilon$-spectrum, which, in turn, causes slower electrons to be overrepresented
in $\partial P_1/\partial \varepsilon$. As the CAP strength decreases, so does this bias. And for a $\gamma_0$-independent spectrum, the singly differential $\partial P_1/\partial \varepsilon$ spectrum may be interpreted as the integrated doubly differential double ionization spectrum,
\begin{equation}
\label{IntepretationDoublyDiff}
\frac{\partial P_1}{\partial \varepsilon} \xrightarrow{\gamma_0 \to 0^+} \int_0^\infty \frac{\partial^2 P_\mathrm{double}}{\partial \varepsilon \partial \varepsilon'} \, d \varepsilon' .
\end{equation}
From this point of view, it is not surprising that the multi-photon peaks seen in $\partial P_1 /\partial \varepsilon$, i.e., the right panels of Figs.~\ref{Fig_ConvergencePhotoElLogScale} and \ref{Fig_ConvergencePhotoOmegaOneLogScale}, are less pronounced than the ones seen in $\partial P_2/ \partial \varepsilon$, i.e. the left panels. This is particularly so for the case in which $\omega=0.3$~a.u. as direct processes are more prominent here than in the case with $\omega=1$~a.u., for which sequential ionization dominates.

While the bias inherent in the sequential nature of the absorption scheme presented here may be undesirable in most situations, it may be of interest in others. For instance, the situation does resemble an experimental situation in the sense that liberated particles are detected one-by-one -- and the most energetic ones first. The similarity between CAPs and detectors, see, e.g., \cite{Kosloff1986,Kvaal2011}, could facilitate comparison with experiment.

It is worthwhile to also address {\it total} ionization probabilities in this context. The converged spectrum obtained from absorbing an electron from the two-particle wave function, $\partial P_2/ \partial \varepsilon$, integrates to the total norm loss from $\Psi_2$. In the case shown in Fig.~\ref{Fig_ConvergencePhotoEl}, Fig.~\ref{Fig_AbsorbTimeGamma0} shows that this probability is close to one. Actually, the integral of $\partial P_2/ \partial \varepsilon$ is slightly less than the total absorption because we in Eq.~(\ref{dPdE_General}) have only projected onto scattering states corresponding to positive (one-particle) energies. This enables us to avoid artificial contributions to the ionization probability from possible populations of Rydberg states which overlap with the CAP.
The ionization probability
\begin{equation}
\label{TotalIonProb}
P_2 = \int_0^\infty \frac{\partial P_2}{\partial \varepsilon} \, d \varepsilon
\end{equation}
is indeed {\it total} in the sense that it includes both single and double ionization; the spectrum
$\partial P_2 / \partial \varepsilon$ is the energy differential probability distribution of the photo electron which is absorbed first -- irrespective of whether also the second electron goes on to be ionized or not. Thus, $\partial P_2 / \partial \varepsilon$ cannot be interpreted as the spectrum of the photo electron emerging from single ionization alone -- unless the probability of double ionization is negligible compared to single ionization. The converged difference between this spectrum and $\partial P_1 / \partial \varepsilon$, however, would correspond to the single ionization event exclusively.

The spectrum $\partial P_1/\partial \varepsilon$ in Eq.~(\ref{dPdE_GeneralOne}) is calculated from the second absorption and, correspondingly, integrates to the double ionization probability.
Note that this quantity,
\begin{equation}
\label{TotalPdouble}
P_1 = \int_0^\infty \frac{\partial P_1}{\partial \varepsilon} \, d \varepsilon ,
\end{equation}
is not subject to the same $\gamma_0$-dependence as is the spectrum $\partial P_1/ \partial \varepsilon$; $P_1$ is the probability of double absorption irrespective of the bias addressed above.
Again, $P_1$
tends to be somewhat lower than the final population of the vacuum state, $p_0(t \rightarrow \infty)$, due to the possible population of Rydberg states which overlap with the CAP. In this respect, it should be mentioned that Rydberg populations in the two-particle system could lead to an undesired population of the one-particle sub-system. Our numerical studies have, however, not shown any indication of this affecting the $\partial P_1 / \partial \varepsilon$-distribution.

\subsection{Concluding remarks}
\label{Sec_ConcludingRemarks}

The original motivation of this work was to enable the description of unbound many-particle systems on truncated numerical grids in a manner which allow us to retain as much information about the system as possible. This is particularly desirable when the outgoing waves span a wide energy region. In such situations, by the time the low-energetic part has left the interaction region, the high-energy part may have travelled quite far -- thus necessitating a very large numerical domain in order to enable a full description. This is the case in, e.g., the in-elastic scattering shown in Fig.~\ref{Fig_ScatteringExample}. Moreover, when doubly excited states are involved, the situation is aggravated further by the fact that parts of the wave packet tends to ``linger'' in the interaction region while high-energy components travel off~\cite{Selsto2013}.

The same feature is seen in the present photo ionization examples; both examples involve both high-energy photo-electrons and photo-electrons with near-zero energy. In the last example, the one with $\omega=1$~a.u. and 10 optical cycles, the full, unabsorbed wave function was more or less contained within a box extending from  $-75$~a.u. to $75$~a.u. at the end of the laser pulse, at $t=T$. With $\gamma_0=2^{-15}$~a.u., the numerical domain had to extend to $\pm 100$~a.u. while $\gamma_0= 2^{-18}$~a.u. required $x$-values up to about $\pm 200$~a.u.
Thus, with the ability to analyze the full wave function right after interaction, this CAP method would not offer much relief in these cases. However, such an ability requires the calculation of projections onto fully correlated scattering states -- either by direct calculations or other indirect means \cite{Palacios2007}. 
In order to avoid this complication, many studies resort to projection onto uncorrelated two-particle scattering states. This is an admissible approach if the wave packet is allowed to propagate further after the laser interaction -- until all of the unbound part of the wave packed reaches the asymptotic region in which the interaction between electrons, bound or unbound, may be neglected. Suppose now that a low-energy photo electron of energy $0.05$~a.u. is to travel into this asymptotic region beyond, say, $x=\pm 100$~a.u. By this time, a three-photon singly-ionized photo electron will have travelled beyond $750$~a.u. Thus, in comparison even the weakest absorber constitutes a significant reduction in the numerical complexity -- despite the fact that CAP-independence must be checked for. With an equidistant numerical grid this particular case corresponds to a reduction to about $(200/750)^2 \approx 7 \%$ of a full-sized two-particle domain. The reduction is even more significant if a wider energy window is to be considered.

As mentioned, a drawback of the method is the fact that only singly-differential spectra are produced. It should also be mentioned that, in the case of photo ionization, the lowest admissible absorber strength is correlated with the pulse duration via the inequality~(\ref{HeissenbergRestrictionBox}); a pulse of long duration $T$ has a narrow bandwidth and, thus, a low $\Delta \varepsilon$ is required. Consequently, also with absorbing boundaries, a larger box is required in order to get correct energy distributions with longer pulses. Nonetheless, these issues does not preclude the present method from facilitating or even enabling the numerical description of several dynamical, unbound many-particle systems.

As absorbing boundary conditions are frequently imposed in simulating the dynamics of unbound two-particle quantum systems, Eq.~(\ref{dPdE_General}) represents a convenient way of extracting relevant information -- information which would have been lost otherwise. There are several situations, such as the one pertaining to Fig.~\ref{Fig_ScatteringExample}, in which the spectrum from the first absorption alone provides relevant and interesting information. And extracting it requires very little additional effort. In addition to solving the time-dependent Schr{\"o}dinger equation, Eq.~(\ref{TDSE}), one simply has to update $\Phi(x,x')$ in Eq.~(\ref{PhiDef}) at each time step and, finally, apply the formula~(\ref{dPdE_General}). The latter involves diagonalization of a simple one-particle operator. Our numerical examples suggest that the resulting spectrum has a rather weak dependence on the characteristics of the CAP function.

If the remainder of the system is to be preserved and spectra corresponding to multiple unbound particles are to be calculated, this also requires the solution of Eq.~(\ref{TDLE_nPart}). 
We have seen that a weaker CAPs may be necessary in order to obtain well resolved energy spectra from the second absorption. We have, however, not given any attention to how the {\it shape} of the CAP function affects the convergence in terms of CAP strength. It would be quite interesting to study whether other choices of CAP functions than Eq.~(\ref{CAPfunction}), or other CAPs than local ones, could provide faster convergence.
This is a topic which merits further investigation. Such an investigation should also aim at formulating precise and general convergence criteria.

In order to accumulate all outgoing waves, the wave must be propagated until even the slowest electrons have reached the CAP region. In the case of photo ionization, this means that the duration of the simulation usually extends considerably beyond the duration of the laser interaction. An interesting question in this regard is wether absorption after explicitly time-dependent interactions could be treated or, at least, facilitated by analytical means, thus evading comparatively time-consuming simulations. In this regard, the works of Refs.~\cite{Palacios2007,Morales2016} are inspirational.
While these issues are beyond the scope of the present work, they will be subject to further investigation.

\section{Conclusions}
\label{Sec_Conclusions}

We have presented an approach to the numerical description of unbound multi-particle quantum systems which allows us to impose absorbing boundary conditions and yet calculate the probability distributions of interest for all of the absorbed particles. In this way we may retain the information of interest about each of the unbound particles while using a numerical domain which is considerably smaller than the actual extension of the wave function. 
This comes about by 
using a complex absorbing potential which, in addition to removing the outgoing, unbound parts of the wave function, also {\it probe} them.
The fact that the absorber is a one-particle operator allows us to analyse the unbound part by projecting onto single-particle scattering states -- as opposed to many-particle scattering states. Consequently, only singly differential spectra are obtained. 

The method lends itself to 
rather straight forward implementation.
It also provides a conceptually appealing 
approach which, via the Lindblad equation, generalizes naturally to any number of particles.

The applicability of the scheme was demonstrated by calculating energy spectra for two examples featuring two-particle models -- one example involving scattering and another involving photo ionization. These calculations demonstrated a rather weak dependence on the strength of the absorbing potential, and the spectra where seen to converge as this strength decreased. In the case of photo ionization, the spectra obtained from the first absorption where seen to converge somewhat faster in absorber strength than the spectra obtained from the second absorption.

\section*{Acknowledgements}

Valuable inputs from dr. Stefanos Carlstr{\"o}m, prof. Sergiy Denysov  and dr. Simen Kvaal are gratefully acknowledged.

\appendix

\section{Derivation of the formulas for differential probabilities}
\label{App_dPdE_TwoPart}

Here we derive the differential probability distribution for absorption from a two and one particle system, Eq.~(\ref{dPdE_General}) and Eq.~(\ref{dPdE_GeneralOne}), respectively, from the more general form of Eq.~(\ref{ExpectationValueRemovedSinglePart}). In doing so, we express both operators, density matrices and state vectors by means of second quantization.
The fermionic two-particle wave function is written
\begin{align}
\label{TwoPartWF}
| \Psi_2 \rangle & = \int \int dx_1 dx_ 2 \, \Psi_2(x_1,x_2) | x_1, x_2 \rangle  = \\
\nonumber
& \frac{1}{\sqrt{2}} \int \int  dx_1 dx_ 2 \, \Psi_2(x_1,x_2) | \{ x_1, x_2 \} \rangle = \\
\nonumber
& \frac{1}{\sqrt{2}} \int \int  dx_1 dx_ 2 \, \Psi_2(x_1,x_2) \hat{\psi}^\dagger(x_1) \hat{\psi}^\dagger(x_2)| - \rangle  ,
\end{align}
where $| x_1 , x_2 \rangle$ is a product basis state, $| \{ x_1 , x_2 \} \rangle$ is a properly anti-symmetrized one and $| - \rangle$ is the vacuum state, i.e., the state in which there are no particles. The two-particle wave function is anti-symmetric with respect to exchange,
\begin{equation}
\label{AsymmPosRepr}
\Psi_2(x_1, x_2) = -\Psi_2(x_2, x_1) .
\end{equation}
We also express the one-particle density matrix and the scattering states by means of second quantization:
\begin{align}
\nonumber
& \rho_1 = \int \int dx dx' \, \rho_1(x, x') | x \rangle \langle x' |  =
\\ &
\label{OnePartDM}
\int \int dx dx' \, \rho_1(x, x') \hat{\psi}^\dagger(x) | - \rangle \langle - | \hat{\psi}(x')
\end{align}
and
\begin{equation}
\label{ScatteringState2ndQuant}
| \varphi_\varepsilon \rangle =
\int dx \, \varphi_\varepsilon(x)
\hat{\psi}^\dagger(x) | - \rangle .
\end{equation}

We start by writing out Eq.~(\ref{ExpectationValueRemovedSinglePart}) for $N=2$ explicitly.
With Eqs.~(\ref{DefGamma}, \ref{TwoPartWF}, \ref{OnePartDM}, \ref{ScatteringState2ndQuant}), it reads
\begin{align*}
&
\hbar \frac{d}{dt} \frac{\partial P}{\partial \varepsilon}=
\int dx  \, \langle \varphi_\varepsilon | \hat{\psi}(x)  \{ \Gamma , | \Psi_2 \rangle \langle \Psi_2 | \}
\hat{\psi}^\dagger(x)| \varphi_\varepsilon \rangle =
\\ &
2 \Re \text{e} \int dx \,
\int dy \int dx' \int \int dx_1 dx_2 \, \varphi_\varepsilon^*(y) \gamma(x') \Psi_2(x_1, x_2)
\\ &
\times \langle - | \hat{\psi}(y) \hat{\psi}(x) \hat{\psi}^\dagger(x') \hat{\psi}(x')
\hat{\psi}^\dagger(x_1) \hat{\psi}^\dagger(x_2)| - \rangle
\\&
\times \int \int dx_1' dx_2' \int dy' \, \Psi_2^*(x_1', x_2') \varphi_\varepsilon(y')
\\ &
\times \langle - | \hat{\psi}(x_2') \hat{\psi}(x_1') \hat{\psi}^\dagger(x) \hat{\psi}^\dagger(y')| - \rangle .
\end{align*}

The vacuum matrix elements may be found, e.g., by using Wick's theorem~\cite{Wick1950}:
\begin{align*}
&
\langle - | \hat{\psi}(y) \hat{\psi}(x) \hat{\psi}^\dagger(x') \hat{\psi}(x')
\hat{\psi}^\dagger(x_1) \hat{\psi}^\dagger(x_2)| - \rangle =
\\ &
\delta(y-x')
\left[\delta(x-x_1)\delta(x'-x_2)- \delta(x-x_2)\delta(x'-x_1)\right]
\\ &
-
\delta(x-x')
\left[\delta(y-x_1)\delta(x'-x_2)-
\delta(y-x_2)\delta(x'-x_1)\right]
\end{align*}
and
\begin{align*}
&
\langle - | \hat{\psi}(x_2') \hat{\psi}(x_1') \hat{\psi}^\dagger(x) \hat{\psi}^\dagger(y')| - \rangle
=
\\ &
\delta(x_2'-y')\delta(x_1'-x) - \delta(x_2'-x)\delta(x_1'-y') .
\end{align*}
With this and the exchange anti-symmetry of the two-particle wave function, Eq.~(\ref{AsymmPosRepr}), we arrive at Eq.~(\ref{dPdETwoPart}).

As explained in Sec.~\ref{Sec_Theory}, when we analyze the part which has been removed from $|\Psi_2 \rangle$, we must make sure to remove the part which is reconstructed within $\rho_1$ in order to avoid double counting.
This contribution is provided by the source term Eq.~(\ref{SourceTerm}). The part to be removed from Eq.~(\ref{dPdETwoPart}) is
\[
\langle \varphi_\varepsilon| \mathcal{S}[\rho_2] | \varphi_\varepsilon \rangle
\]
with $\rho_2 = | \Psi_2 \rangle \langle \Psi_2 |$.
Using Eqs.~(\ref{SourceTerm}, \ref{TwoPartWF}, \ref{ScatteringState2ndQuant}) it may be expressed as
\begin{align*}
&
2 \langle \varphi_\varepsilon| \int dx \, \gamma(x) \hat{\psi}(x) | \Psi_2 \rangle \langle \Psi_2 | \hat{\psi}^\dagger(x)
| \varphi_\varepsilon \rangle
=
\\ &
\int dy \, \varphi_\varepsilon(y) \int dx \, \gamma(x) \int \int dx_1 dx_2 \Psi_2(x_1, x_2 )
\\ &
\times
\langle - | \hat{\psi}(y) \hat{\psi}(x)
\hat{\psi}^\dagger(x_1) \hat{\psi}^\dagger(x_2) | - \rangle
\\ &
\times \int dy' \, \int \int dx_1' dx_2' \, \varphi_\varepsilon(y') \Psi_2^*(x_1', x_2')
\\ &
\times
\langle - | \hat{\psi}(x_2') \hat{\psi}(x_1') \hat{\psi}^\dagger(x) \hat{\psi}^\dagger(y') | - \rangle .
\end{align*}
The repeated vacuum matrix element is
\begin{align*}
& \langle - | \hat{\psi}(y) \hat{\psi}(x)
\hat{\psi}^\dagger(x_1) \hat{\psi}^\dagger(x_2) | - \rangle
=
\\ &
\delta(y-x_2)\delta(x-x_1) - \delta(y-x_1) \delta(x'-x_2) .
\end{align*}
With this and Eq.~(\ref{AsymmPosRepr})
we arrive at
\begin{align*}
& \langle \varphi_\varepsilon| \mathcal{S}[\rho_2] | \varphi_\varepsilon \rangle =
\\ &
\int dx \, \gamma(x) \int dy \, \varphi^*_\varepsilon(y)  2 \Psi_2(x, y) \int dy' \, \varphi_\varepsilon(y) 2 \Psi_2^*(x,y') =
\\ &
4 \int dx \, \gamma(x) \left| \int dy \, \varphi^*_\varepsilon(y)  \Psi_2(x, y) \right|^2 .
\end{align*}
This coincides the last term in Eq.~(\ref{dPdETwoPart}), which, accordingly, is to be removed.

\section{A propagator for the one-particle density matrix}
\label{App_Propagator}

A second order Taylor expansion of $\rho_1$ in time says that
\begin{equation}
\label{Taylor}
\rho_1(t+\tau) = \rho_1 + \tau \dot{\rho}_1 + \frac{1}{2} \tau^2 \ddot{\rho}_1 + \mathcal{O}(\tau^3) .
\end{equation}
Here, the dots indicate time-derivatives and, for convenience, the absence of an argument is to be interpreted as ``$(t)$''. $\dot{\rho}_1$ and $\ddot{\rho}_1$ are provided by Eq.~(\ref{TDLE_1part}) and its time derivative, respectively. If we write them out explicitly, Eq.~(\ref{Taylor}) reads
\begin{align}
\nonumber
&\rho_1(t+\tau) = \rho_1
-i \frac{\tau}{\hbar} \left( h_\mathrm{eff} \rho_1 - \rho_1 h_\mathrm{eff}^\dagger + i \mathcal{S}[\Psi_2] \right)
\\ &
\nonumber
+ \frac{\tau^2}{2 \hbar^2} \bigg( \left(-i \hbar \, \dot{h}_\mathrm{eff}- h_\mathrm{eff}^2 \right) \rho_1 -
\\ &
\nonumber
\rho_1 \left( -i \hbar \, \dot{h}_\mathrm{eff}^\dagger+(h_\mathrm{eff}^\dagger)^2 \right)
\\ &
\nonumber
+ 2 h_\mathrm{eff} \rho_1 h_\mathrm{eff}^\dagger - i \left[ h_\mathrm{eff} \mathcal{S}[\Psi_2] - \mathcal{S}[\Psi_2] h_\mathrm{eff}^\dagger \right]
\\ &
\label{Explicit2ndOrder}
+ \hbar \frac{d}{dt}\mathcal{S}[\Psi_2]
\bigg)   + \mathcal{O}(\tau^3) ,
\end{align}
where $h_\mathrm{eff}$ is here the effective one-particle Hamiltonian.
In an autonomous system, $\dot{h}_\mathrm{eff}$ vanishes and the scheme is somewhat simplified.

Now, the sum of the terms which do not contain source term contributions may, to third order in $\tau$, be written as
\begin{equation}
\label{MagnusForMaster}
\exp\left(-i \frac{\tau}{\hbar} \, h_\mathrm{eff}(t+\tau/2) \right)
\rho_1
\exp\left(i \frac{\tau}{\hbar} \, h_\mathrm{eff}^\dagger(t+\tau/2) \right) .
\end{equation}
Moreover,
\begin{align}
\label{SourceTermProp}
& \frac{\tau}{\hbar} \mathcal{S}[\Psi_2] + \frac{\tau^2}{2 \hbar} \frac{d}{dt} \mathcal{S}[\Psi_2] =
\\ &
\nonumber
\frac{\tau}{2\hbar} \left( \mathcal{S}[\Psi_2] + \mathcal{S}[\Psi_2(t+\tau)]\right) + \mathcal{O}(\tau^3) ,
\end{align}
which allows for a convenient implementation simply by keeping the previous source term in memory.

All in all, we arrive at the following scheme:
\begin{align}
\nonumber
& \rho_1(t+\tau) = e^{-i \tau/\hbar \, h_\mathrm{eff}(t+\tau/2)}
\rho_1(t)
e^{i \tau/\hbar \, h_\mathrm{eff}^\dagger(t+\tau/2)}
\\ &
\label{Rho1Scheme}
+ \frac{\tau}{2\hbar} \left( \mathcal{S}[\Psi_2(t)] + \mathcal{S}[\Psi_2(t+\tau)]\right)
\\ &
\nonumber
- i \frac{\tau^2}{2 \hbar^2} \left[ h_\mathrm{eff}(t) \mathcal{S}[\Psi_2(t)] -
\mathcal{S}[\Psi_2(t)] h_\mathrm{eff}^\dagger(t)]\right] + \mathcal{O}(\tau^3) .
\end{align}

\end{document}